\documentclass[a4paper, 12pt]{article}

\usepackage[sort&compress]{natbib}
\bibpunct{(}{)}{;}{a}{}{,} 

\usepackage{amsthm, amsmath, amssymb, mathrsfs, multirow, url, subfigure}
\usepackage{graphicx} 
\usepackage{ifthen} 
\usepackage{amsfonts}
\usepackage[usenames]{color}
\usepackage{fullpage}
\usepackage{tikz}
\usepackage{bm}

\theoremstyle{plain}

\theoremstyle{definition}

\theoremstyle{remark}

\newcommand{\E}{\mathsf{E}}

\newcommand{\unif}{{\sf Unif}}
\newcommand{\nm}{{\sf N}}

\newcommand{\stt}{{\sf t}}

\title{Incorporating Asymmetric Loss for Real Estate Prediction with Area-level Spatial Data}
\author{Vaidehi Dixit\footnote{Department of Statistics, University of Missouri; {\tt vdixit@missouri.edu}, {\tt holans@missouri.edu}, {\tt wiklec@missouri.edu}}, \quad Scott H. Holan$^{*}$\footnote{Research and Methodology Directorate, US Census Bureau}, \quad and \quad Christopher K. Wikle$^*$}
\date{\today}

\begin{document}

\maketitle 

\begin{abstract}
We investigate two asymmetric loss functions, namely LINEX loss and power divergence loss for optimal spatial prediction with area-level data. With our motivation arising from the real estate industry, namely in real estate valuation, we use the Zillow Home Value Index (ZHVI) for county-level values to show the change in prediction when the loss is different (asymmetric) from a traditional squared error loss (symmetric) function. Additionally, we discuss the importance of choosing the asymmetry parameter, and propose a solution to this choice for a general asymmetric loss function. Since the focus is on area-level data predictions, we propose the methodology in the context of conditionally autoregressive (CAR) models. We conclude that choice of the loss functions for spatial area-level predictions can play a crucial role, and is heavily driven by the choice of parameters in the respective loss.

\smallskip

\end{abstract}

\section{Introduction}
The use of asymmetric loss functions is abundant in the literature and yet their full potential remains unexplored. In this paper, we explore asymmetric loss functions with respect to decision-making in the real estate industry, namely with respect to real estate valuation using area-level spatial data. The default loss function used in most analyses is squared error loss. This assumes equal loss for the same size of overestimation and underestimation. The need for asymmetric loss arises because the aforementioned costs could be uneven. For example, if a commodity is being sold and the price is underestimated, it would quickly get sold but incur a loss for the seller. On the other hand, if the valuation is too high, it may not get sold at all, and the seller still experiences a loss. The loss incurred in either direction could be different and needs to be taken into account when deciding the price. 

A commonly used asymmetric loss is the LINEX loss, which has a linear and exponential term creating the asymmetry. It has been used for both estimation and prediction problems, incorporated in both frequentist and Bayesian methodologies. Introduced in \citet{varian1975bayesian} for real estate assessment, optimal predictors under LINEX loss have been developed in studies like \citet{zellner1986bayesian}, \citet{christoffersen1997optimal} and \citet{basu1991bayesian}. Effects of the LINEX loss have been investigated for forecasting in economics studies like \citet{patton2007properties}, \citet{capistran2009disagreement} and \citet{elliott2008biases}, to name a few. The other asymmetric loss function we focus on in this paper is newer and is known as the power divergence loss (PDL), which was introduced in \citet{cressie2022optimal} in the context of spatial prediction. The authors point out that although the notion of asymmetry in loss is often required in spatial problems, it has rarely been considered, partially due to the convenience of the squared error loss function, as it leads to unbiased estimators. This convenience is also reflected in the kriging predictor, which is based on the squared error loss. With the need for asymmetric loss in spatial literature emphasized in \citet{cressie2023decisions} and \citet{cressie2022optimal}, \citet{pearse2024optimal} developed optimal spatial prediction under power divergence loss. A unique feature of this loss function over the LINEX loss is that it produces non-negative predictions for such observed data. Given that this loss function is still underexplored in spatial literature, we aim to investigate it, in addition to the LINEX loss for spatial prediction of area-level data.

Both the LINEX loss and PDL  depend on the choice of an asymmetry parameter. This paper also explores this aspect. The approaches suggested in literature to select this parameter have been subjective. In \citet{cain1995}, the authors use the elicitation of a domain expert to choose this parameter or minimize expected LINEX loss over a grid of values. The former can be an excellent strategy if the relevant domain knowledge is available. In \citet{cressie2022optimal}, the authors propose a quantile matching method, such that the resulting prediction matches a specified quantile. This can be helpful in terms of comparing predictions based on quantiles and that with asymmetric loss. Further, \citet{jayasinghe2022regression} constructed a {\em power ratio} (PR) that aims to balance the need for asymmetry while guarding against too much asymmetry. The recommendation in this paper was in the context of a regression fit using LINEX loss. We explore this approach in our paper for spatial predictions using area-level data.

Area-level data are collected for diverse applications, each with different objectives. For example in public health and epidemiology, disease counts are available by county and health authorities can identify areas of risk and plan interventions accordingly. Asymmetric loss is important in socioeconomic studies, where policymakers can identify social irregularities using aggregated social variables across regions, or better economic policies can be formulated with area-level data on commodities. Such loss functions also play a significant role in environmental monitoring, where climate scientists obtain area-level data in the form of ecodiversity and associated climate variables, allowing for effective decisions concerning climate change and sustainable development. In all these situations, decision-making is crucial, and the costs associated with these decisions can be substantial. In most cases, the costs associated with overestimation or underestimation are not the same. For instance, in the case of a flood or spread of infectious disease, the costs of mitigation are often much lower than the costs of damage. This necessitates using an asymmetric risk associated with the corresponding decisions. Since our focus is on area-level data, we present our optimal spatial predictions under a general Bayesian hierarchical framework \citep[e.g.,][]{banerjee2003, cressiewikle2011} for area-level data. The hierarchy represents a data (observed) level, a process (latent) level, and a parameter level. The process level of this framework characterizes the spatial nature of the problem, enabling the user to incorporate any spatial dependencies. In particular we illustrate the use of a conditionally autoregressive (CAR) model at the process level, that models the mean of each region as a function of its neighboring regions. This is commonly used to model disease counts in epidemiology studies, or for survey analysis with region-wise counts. 

Here, our example is concerned with real estate valuation. Understanding real estate valuation requires incorporating spatial dependence and recognizing the asymmetry in overestimation and underestimation. In fact, the LINEX loss function was introduced in the context of real estate assessment by \citet{varian1975bayesian}. Subsequent work incorporating LINEX in real estate has appeared in \citet{cain1995} and \citet{parsian2002estimation}. Our goal in this paper is to combine the concepts of spatial prediction and asymmetric loss in assessing real estate valuation using area-level data.

The remainder of this paper proceeds as follows. In Section \ref{s:motivation_methodology} we first describe our motivating real estate dataset, followed by the methodological features. In the latter we review optimal prediction under the two loss functions -- LINEX and PDL, and propose the hierarchical spatial model for area-level data. In Section \ref{s:analysis} we illustrate this methodology for the motivating dataset, while comparing the optimal predictions under asymmetric loss with the squared error loss. The paper concludes with a discussion on future work in Section \ref{s:discussion}.

\section{Motivation and methodology}
\label{s:motivation_methodology}
Before we describe the methodology for optimal prediction in area-level data using asymmetric loss, we first describe our motivating dataset. This comprises real estate valuations that are county-level typical values of a home in the state of New Jersey. Motivated by these area-level data, we split our methodology in three parts. We first describe a data and process model for area-level data, then we review the two loss functions, and finally present a particular CAR model for modeling area-level data. 
\subsection{Real estate industry dataset}
\label{s:dataset}
\begin{figure}
\centering
\subfigure[]
{\includegraphics[width = 0.38\textwidth]{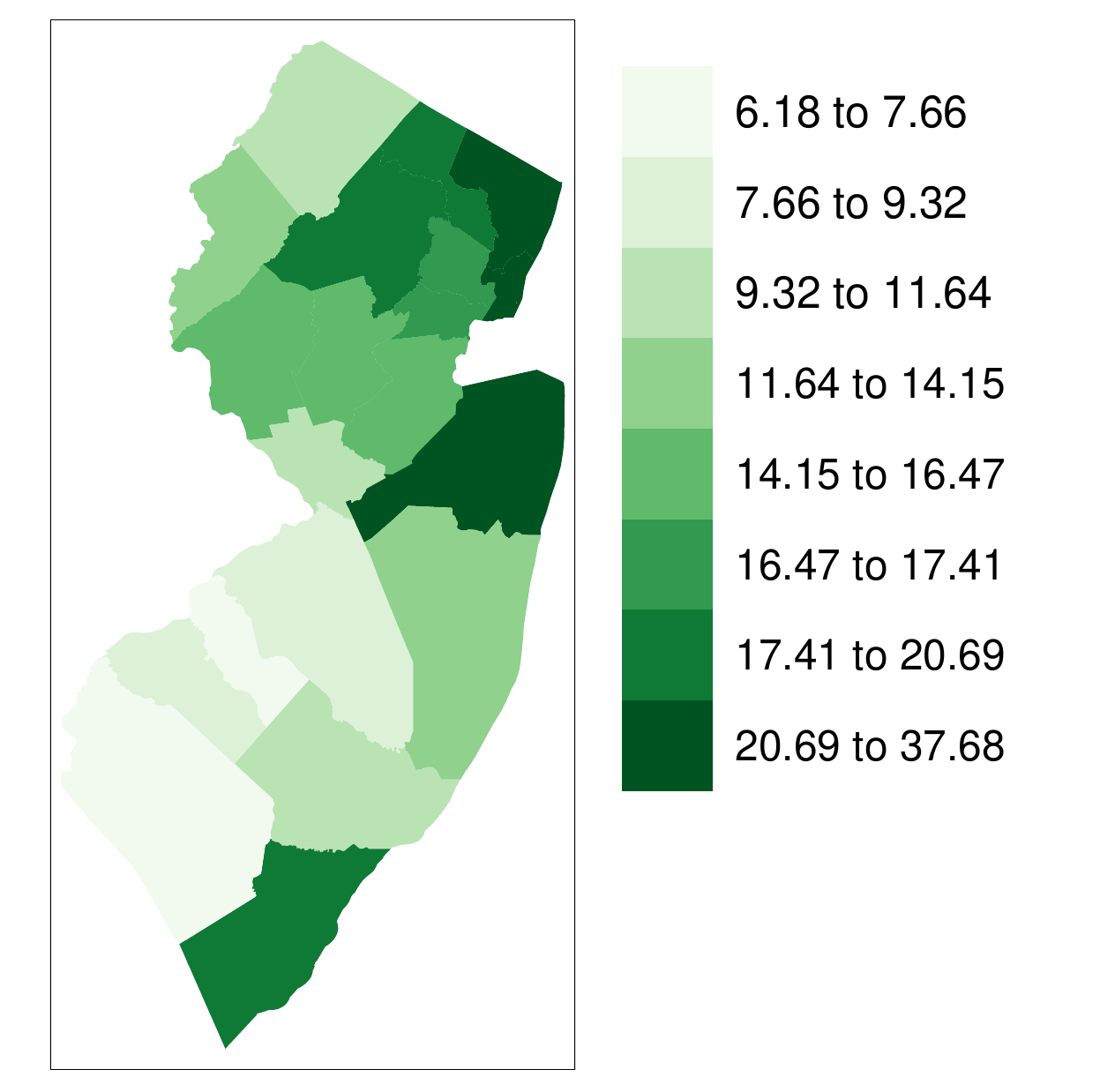}}
\subfigure[]
{\includegraphics[width = 0.39\textwidth]{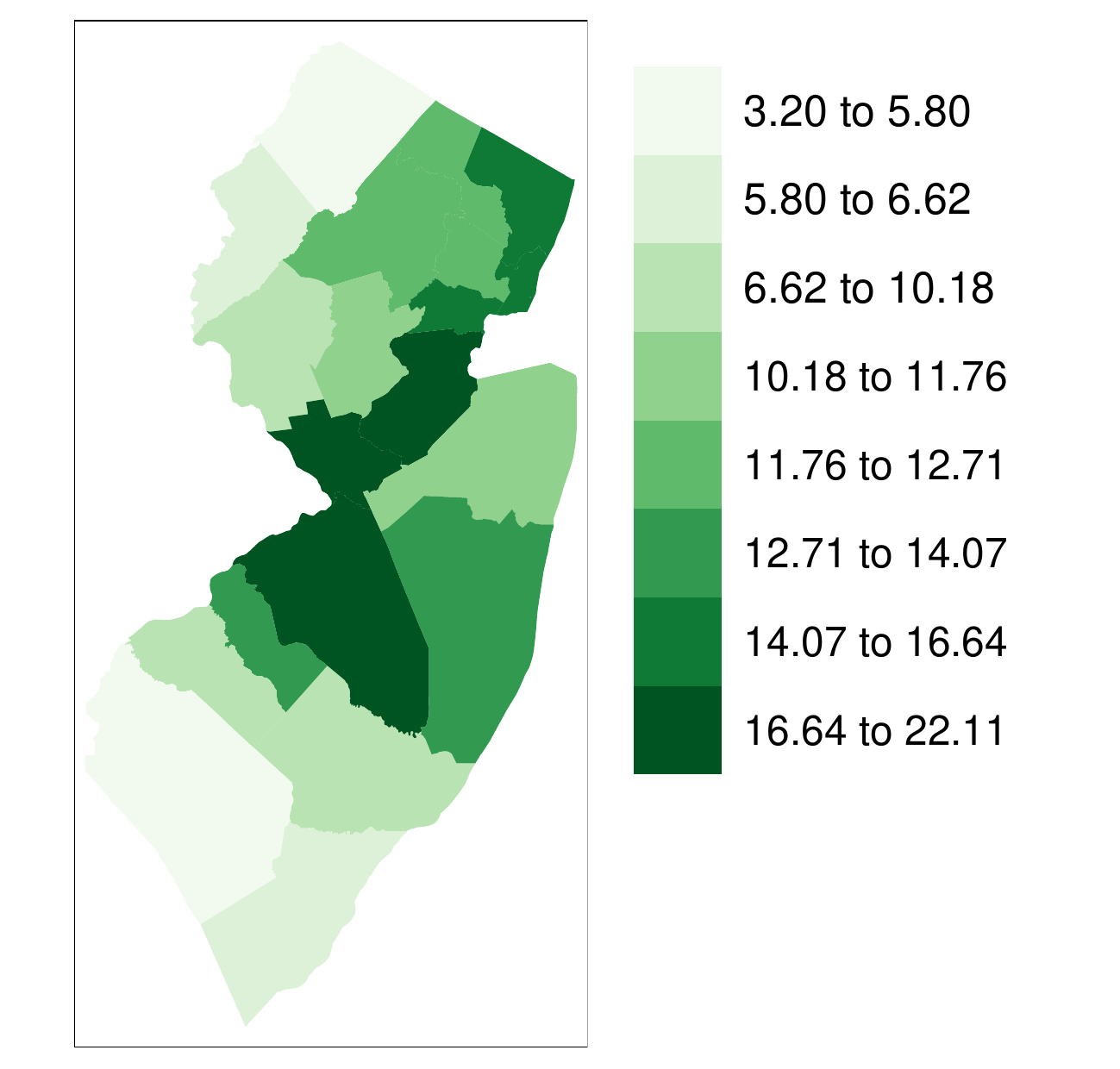}}
\subfigure[]
{\includegraphics[width = 0.38\textwidth]{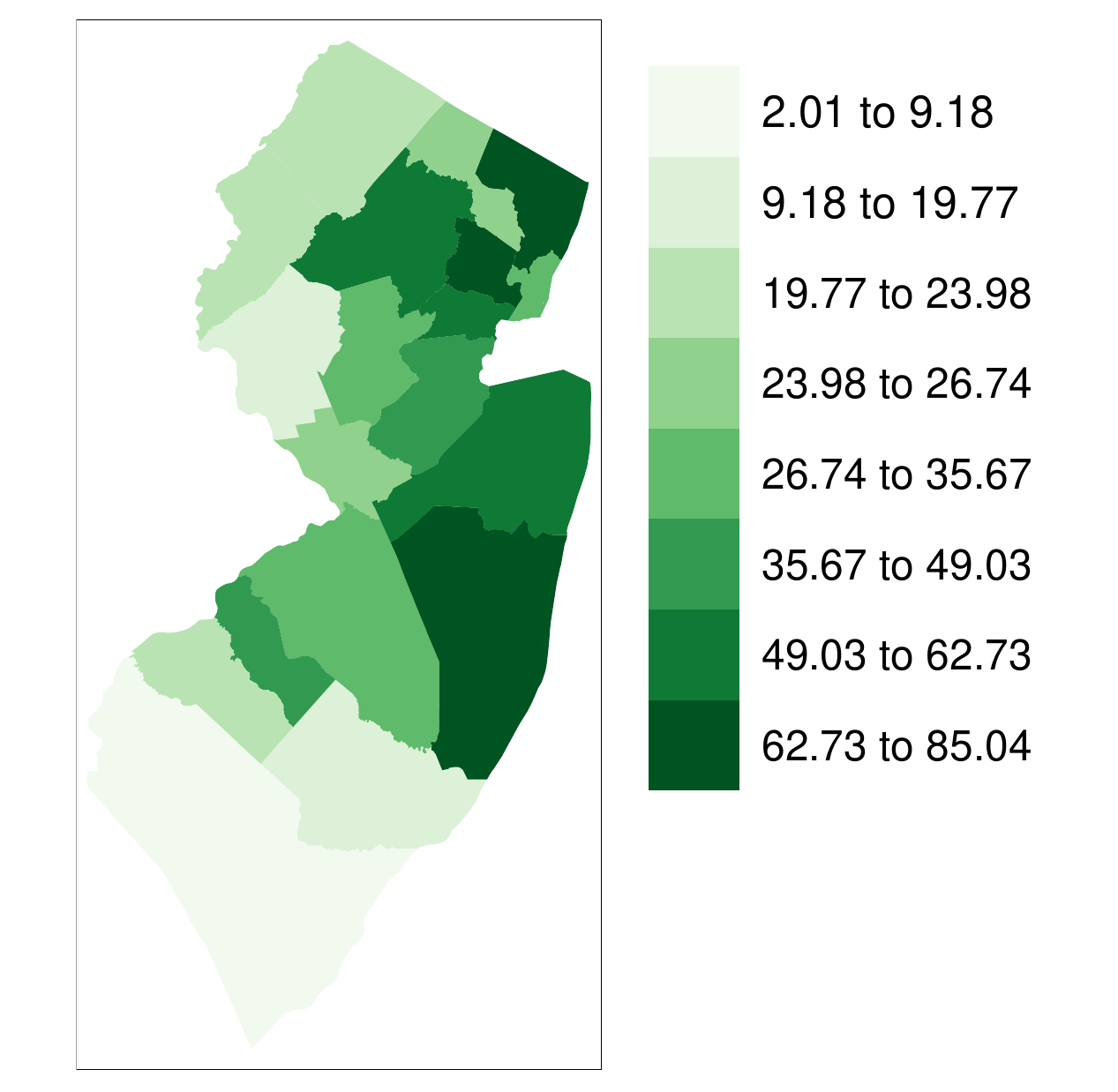}}
\subfigure[  ]
{\includegraphics[width = 0.4\textwidth]{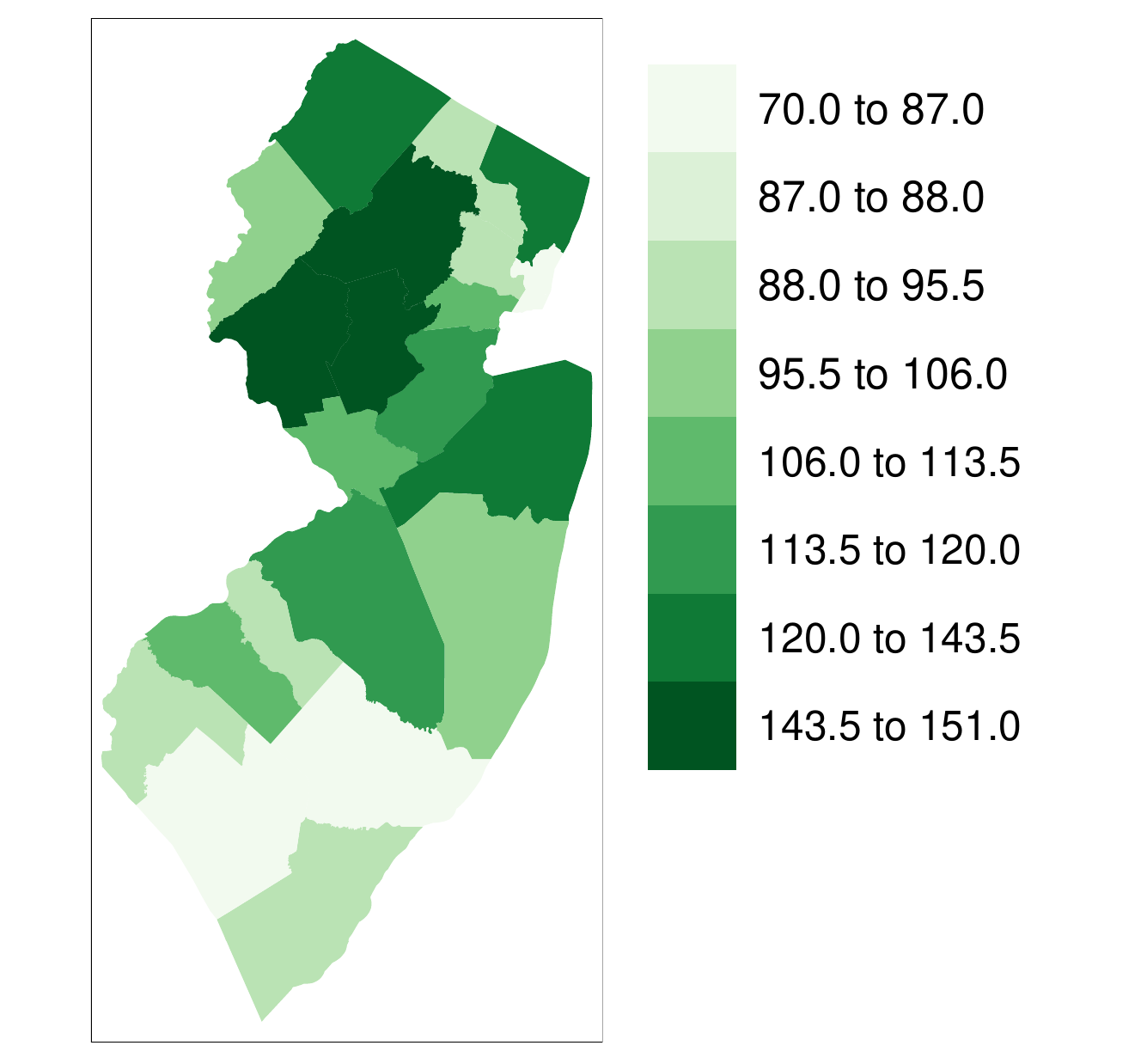}}
\caption{County-level values of (a) Zillow Home Value Index (ZHVI) values (in dollars) for the month of April in 2018 in NJ scaled by $10^4$, (b) the number of movers in county-to-county flow for the year 2018 in NJ scaled by $100$, (c) the number of construction permits authorized in year 2018 in NJ scaled by $100$, (d) the median household income in 2017-2021 in NJ scaled by $1000$.}
\label{fig:firstdata}
\end{figure}

The dataset we use is available for download on Zillow \footnote{https://www.zillow.com/research/data/}. Zillow constructs an index called ZHVI (Zillow Home Value Index) that is based on all homes in the particular region (e.g. county/state). This has been used in economic studies like \citet{glynn2022learning} and \citet{holt2020using}. This index not only accounts for the individual attributes of the house but also a history of sales transactions, tax assessments, and public records from the respective jurisdiction. The ZHVI is aimed at estimating the price of a home in the $35^{th}$ to $65^{th}$ percentile. These are available across various home sizes as region-wise estimates. We focus on the typical county-level home value of a 3-bedroom home in the state of New Jersey (NJ) for all 21 counties. The home prices in New Jersey have large spatial variance given that NJ has major cities (New York city and Philadelphia) on the east and west end of the state. This index is available as a monthly value available for all years 2000-2024. Additionally, we consider three covariates in our model. Specifically, we use the county-level migration 5-year period estimates for years 2016-2020 obtained from the American Community Survey (ACS)\footnote{https://www.census.gov/topics/population/migration/guidance/county-to-county-migration-flows/2016-2020.html}, the number of construction permits authorized in NJ for the year obtained from the NJOIT open data center\footnote{https://data.nj.gov/Reference-Data/Housing-Units-Authorized-by-Building-Permits/v5ve-akg6} and the ACS 5-year period estimate for the county-level median household income\footnote{HDPulse: An Ecosystem of Minority Health and Health Disparities Resources. National Institute on Minority Health and Health Disparities. Created 6/24/2024. Available from https://hdpulse.nimhd.nih.gov} for 2017-2021. Since the covariate of migration is available as a 5-year period estimate for 2016-2020, we focus on the pre-COVID year 2018 and use the ZHVI index in the month of April as most sales happen in the Spring season. Area-level plots of the ZHVI index and the three covariates are given in Figure \ref{fig:firstdata}.

\subsection{Methodology}
\label{s:methodology}
We first describe a general Bayesian hierarchical model for area-level data followed by a review of the loss functions and a particular CAR model.
\subsubsection{Data and process model}
\label{s:data_process_model}
As the focus of this paper is area-level data, we consider a general hierarchical model for area-level data as found in spatial literature like \citet{banerjee2003} and \citet{cressiewikle2011}. The model is hierarchical in the sense that we have a conditional model of the data given the underlying process \citep[see][]{cressiewikle2011, wikle2003hierarchical}. Let us suppose our spatial domain is divided into $n$ spatial areas and we observe data $\bm Z = (Z_1, \ldots, Z_n)^\intercal$ corresponding to the latent process vector $\bm Y = (Y_1, \ldots, Y_n)^\intercal$. Then the data model of the hierarchy is given by
\[Z_i = Y_i + \varepsilon_i, \quad i=1, \ldots, n,\]
where $\varepsilon_i$ is the random measurement error associated with the $i^{th}$ region. The underlying process $Y_i$ is then modeled as
\[Y_i = \mu_i + \eta_i, \quad i=1, \ldots, n,\]
where $\mu_i$ represents the mean of the process at the $i^{th}$ region and can be modeled with area-level covariates $\bm X_i$ as $\mu_i = \bm X_i^\intercal \bm \beta$. The random component $\eta_i$ characterizes the spatial dependence. We assume $\varepsilon_i$ is independent of $\eta_i$ and 
$\varepsilon_i \sim \text{Gau}(0, \sigma^2 \bm I_n)$ where $\sigma^2$ denotes the measurement error variance and can often be assumed known. The spatial component $\eta_i$ is assumed to be $\text{Gau} (0, \bm H_n (\bm \gamma))$ where $\bm H_n$ is a $n \times n$ variance-covariance matrix accounting for spatial adjacencies and is a function of the vector of hyperparameters $\bm \gamma$. 

\noindent The data model conditional on the underlying latent process can then be written as
\[Z_i | Y_i \sim \text{Gau}(\mu_i, \sigma^2 \bm I_n + \bm H_n (\gamma)), \quad i=1, \dots, n.\]
We illustrate a case of the process model that assigns a structure to the spatial dependence in the following subsection.


\subsubsection{CAR model}
\label{ss:CAR}
We now present a particular case of the process model, i.e., Gaussian CAR models. Specifically, the mean for each $Y_i$ conditional on the other values is modeled as a function of its neighboring regions, i.e,
\[\E(Y_i \mid Y_j, j\neq i) =  \mu_i + \rho \sum_{j=1}^n c_{ij} (y_j - \mu_j),\]
which can also be written as
\[\E(Y_i \mid Y_j, j\neq i) =  \mu_i + \phi_i,\] 
where $\phi_i = \rho \sum_{j=1}^n c_{ij} (y_j - \mu_j)$ is the spatial trend and $c_{ij}$ is the $(i,j)^{th}$ element of the adjacency matrix $C$.
This gives the joint distribution of $\bm Y = (Y_1, \ldots, Y_n)^\intercal$ conditional on the parameters $(\bm \mu, \rho, \tau_1, \ldots, \tau_n)$ as
\[\mathbf{Y} \sim \nm (\bm{\mu}, [\mathbf{I} - \rho \mathbf{C}]^{-1} \mathbf{M}),\]
where $\bm{\mu}$ models the joint mean of the process, $\rho \in (-1,1)$ is the autocorrelation parameter, $\bm C$ is the adjacency matrix and $\bm M$ is a diagonal matrix of variances. The adjacency matrix $\bm C$ is known such that $c_{ii} = 0 \quad \forall~ i$ and $c_{ij} \neq 0$ only if $j$ is in the neighbourhood of $i$. Matrix $\bm M$ can be expressed as $\tau_i^2 \bm I_n$ where $\tau_i^2$ is the variance parameter for region $i$. If $\tau_i^2 = \tau^2$ for each $i$ then it is assumed that the regions have equal variance (see \citet{banerjee2003} and \citet{cressiewikle2011} for further discussion).
The mean vector $\bm \mu$ can be expressed in terms of an intercept and $p$ covariates as
\[\mu_i = \beta_0 + \sum_{j=1}^{p} X_{ij} \beta_j, \quad i=1, \ldots, n.\]
For a Bayesian implementation of the CAR model, priors are assigned to each of the parameters $\rho$, $\tau$, and the regression coefficients $\bm \beta$. Posterior distributions can then be constructed for each parameter and posterior draws can be obtained. Note that the posterior draws for the fitted values will include the spatial trend $\phi$. Suppose we obtain $M$ draws $\hat Y_{i}^{(1)}, \ldots, \hat Y_{i}^{(M)}$ for each of the regions $i=1, \ldots, n$. The Bayesian approach is to take the posterior mean of the  $M$ draws to obtain a prediction for each region $i \in \{1, \ldots, n\}$. However, the posterior mean is the optimal predictor under the squared error loss function and hence assumes symmetric loss. At this juncture, we employ the use of an asymmetric loss function
and calculate the predictions based on the optimal predictors in the subsequent section. 



We now review the two asymmetric loss functions and obtain the optimal predictors under both for the above model.


\subsubsection{Review of loss functions}
\label{s:reviewloss}
Incorporation of asymmetric loss can happen at any stage of analysis. From a frequentist perspective, replacing squared error loss with an asymmetric loss in a regression model can directly impact the coefficient estimates. In a Bayesian setting, in addition to the likelihood stage, the asymmetric loss can be implemented at the prediction stage, i.e. instead of using a posterior mean that minimizes the squared error loss, the posterior draws can be summarized with the optimal predictor under a different loss function. For spatial area-level data, this can be interpreted in terms of conditional autoregressive models, where the posterior predictions are calculated by taking a posterior mean. We focus on the use of an asymmetric loss function at the fitting/prediction stage of a Bayesian model for area-level data. This means that instead of using the posterior mean, the optimal predictor will be determined based on the asymmetric loss function. With that in mind, we define our notation as follows. Let $\bm Y = (Y_1, Y_2, \ldots, Y_n)^\intercal$ be the latent process values for area $i$ and $\hat {\bm Y} = (\hat Y_1, \hat Y_2, \ldots, \hat Y_n)^\intercal$ be the corresponding predictions based on the posterior distribution of the process given the data and $\Delta_i = \hat Y_i -  Y_i$ be the prediction error for the $i^{th}$ spatial region. We first review how these are obtained under both the LINEX and power divergence loss (PDL) below.

\begin{itemize}
\item \textbf{LINEX loss}

The LINEX loss function for prediction for $i^{th}$ region is given by
\begin{equation}
\label{eq:linexloss}
    L(\Delta_i) = \gamma \Bigl [ \exp{(\lambda \Delta_i)} - \lambda \Delta_i - 1\Bigr ],
\end{equation}
where $\lambda \neq 0$ is the shape parameter and $\gamma > 0$ is the scale parameter. When $\lambda <0$, under estimation is more costly as the exponential term will grow for negative $\Delta_i$ and when $\lambda>0$, over estimation is more costly than  under estimation. Since the role of $\gamma$ is only to scale the loss function, it does not affect the asymmetry of the loss. For joint prediction of $n$ observations, the joint LINEX loss is a sum over $i$ in \eqref{eq:linexloss}. Note that this is equivalent to the extended LINEX loss in \citet{zellner1986bayesian} with $\lambda_i = \lambda ~\forall i \in \{1, \ldots, n\}$.

To obtain the optimal predictor under LINEX loss, we have to consider the expected loss under the posterior distribution of $\bm Y | \bm Z$, i.e.,
\[\E_{\bm Y | \bm Z}(L(\Delta)) = \gamma \sum_{i=1}^n \Bigl[\exp(\lambda \hat Y_i).\E_{Y_i | Z_i} [\exp(-\lambda Y_i)] - \lambda \hat Y_i + \lambda \E_{Y_i | Z_i} [Y_i] - 1 \Bigr ].\]
The estimator that minimizes this loss is given by \citet{zellner1986bayesian} as
\[\hat Y_{i}^{\text{LINEX}} = (-1/\lambda) \log \E_{Y_i \mid Z_i}(\exp {(-\lambda Y_i)}) \quad \forall~ i \in \{1, \ldots, n\}.\]
Calculation of this optimal estimator is non-trivial. For a conjugate Bayesian model, where the posterior distribution is Gaussian, an analytical form of this estimator has been obtained in \citet{zellner1986bayesian}. In general this can be reasonably approximated by Monte Carlo techniques, based on $M$ draws from the posterior distribution,
\begin{equation}
\label{eq:linexestimate}
    \hat Y^{\text{LINEX}}_i = -\frac{1}{\lambda} \log \left(\frac{1}{M}\sum_{j=1}^{M} \exp{(-\lambda \hat Y_{i}^{(j)})}\right).
\end{equation}

The value of the shape parameter $\lambda$ can heavily influence the direction and magnitude of the asymmetry of this loss. In the literature this has been dealt with using elicitation from domain experts, minimizing expected LINEX loss over a grid of values \citep{cain1995}, or pre-determining reasonable values \citep{varian1975bayesian}. Even though reasonable in the particular context, this does not guide a user in using the LINEX loss function with appropriate $\lambda$ for other purposes and remains a point of discussion. \citet{jayasinghe2022regression} identified this gap and suggest a power ratio function that helps determine an optimum shape parameter in a means similar to that of a scree plot in the principle component analysis. We take their suggestion and implement it for both, LINEX loss and PDL as described below.

\item \textbf{Power divergence loss}

Introduced in \citet{cressie2022optimal}, a thorough study on the power divergence loss (PDL) function has been conducted in \citet{pearse2024optimal}. The main motivation is to develop a loss function based on the phi-divergence that is appropriate when one is predicting non-negative spatial processes. Like the LINEX loss, the asymmetry in the power divergence loss is characterized by a shape parameter $\lambda \in (-\infty, \infty)$. The power divergence loss for the prediction for spatial region $i$ is given by
\begin{equation}
\label{eq:pdlloss}
\begin{aligned}
    L(\Delta_i) &= \frac{1}{\lambda(\lambda + 1)} \left [ Y_i \left(\left(\frac{Y_i}{\hat Y_i}\right)^\lambda - 1\right) + \lambda(\hat Y_i - Y_i) \right ], \quad \lambda \neq 0, -1,\\
    &= Y_i \log {\frac{Y_i}{\hat Y_i}} - (Y_i - \hat Y_i), \quad \lambda = 0,\\
    &= (Y_i - \hat Y_i) - \hat Y_i \log {\frac{Y_i}{\hat Y_i}}, \quad \lambda = -1.
\end{aligned}
\end{equation}
This loss is a continuous and convex function of $(Y_i, \hat Y_i)$ for $Y_i, \hat Y_i >0$. Similar to the LINEX loss estimator, we will consider the optimal predictor under the expected power divergence loss  with respect to the posterior distribution of $\bm Y| \bm Z$, which is
\begin{equation}
\label{eq:pdlrisk}
\begin{aligned}
    \E_{\bm Y | \bm Z}(L(\Delta)) &= \frac{1}{\lambda(\lambda + 1)} \sum_{i=1}^n \left [ \E_{Y_i | Z_i}  \left \{Y_i \left(\left(\frac{Y_i}{\hat Y_i}\right)^\lambda - 1\right) \right \} + \lambda \hat Y_i - \lambda \E_{Y_i | Z_i} Y_i \right ], \quad \lambda \neq 0, -1,\\
    &= \sum_{i=1}^n \left [\E_{Y_i | Z_i} \left \{Y_i \log {\frac{Y_i}{\hat Y_i}} \right \} - \E_{Y_i | Z_i} Y_i + \hat Y_i\right ], \quad \lambda = 0,\\
    &= \sum_{i=1}^n \left [ \E_{Y_i | Z_i} Y_i - \hat Y_i - \hat Y_i \, \E_{Y_i | Z_i} \log \frac{Y_i}{\hat Y_i}   \right], \quad \lambda = -1.
\end{aligned}
\end{equation}

The estimator that minimizes this loss 
$\E_{\bm Y | \bm Z}(L(\Delta))$ is given by \cite{pearse2024optimal} as
\begin{equation}
\label{eq:pdl.predictor}
\begin{aligned}
    \hat Y_i^{PDL} &= (\E_{Y_i | Z_i} (Y_i^{\lambda + 1}))^{1/(\lambda + 1)}, \quad \lambda \neq -1,\\
    &= \exp \{\E_{Y_i | Z_i} (\log Y_i)\}, \quad \lambda = -1, 
\end{aligned}
\end{equation}
whose MCMC approximation is,
\begin{equation}
\begin{aligned}
\label{eq:pdlestimate}
     \hat Y^{\text{PDL}}_i &= \left (\frac{1}{M} \sum_{j=1}^M \hat Y_{i}^{(j)(\lambda + 1)}\right)^{1/(\lambda + 1)}, \quad \lambda \neq -1,\\
      &= \exp\left \{\frac{1}{M} \sum_{j=1}^{M} \log {(\hat Y_{i}^{(j)})} \right\}, \quad \lambda = -1.
\end{aligned}
\end{equation}



Here again the choice of the parameter $\lambda$ plays an important role in determining the degree and direction of asymmetry of the loss. In \citet{pearse2024optimal}, the authors suggest calibrating the optimal PDL estimator in terms of $\lambda$ such that it estimates a specified quantile of the posterior distribution. Based on the quantile the user chooses, the resulting estimator under PDL will over estimate or under estimate the required quantity as compared to the squared error loss. However, the interpretation of the respective PDL loss on its own remains incomplete, as the optimal predictor is not achieved via a pre-specified, needed asymmetry in the loss, but the optimal predictor (quantile) points to the asymmetry achieved.

\item \textbf{Choosing the asymmetry parameter}
\label{ss:asypar}

The asymmetry parameter in either of the loss functions above is important, as it decides the magnitude and direction of the loss. At the same time, determining the appropriate value of this parameter is a challenging task. A very natural first approach is to minimize the expected loss over a grid of the parameter values. However, this does not serve the right purpose. As an example, consider the case where over estimation is more costly and hence under estimation is desired. In the case of the LINEX loss, this means that the appropriate value of the shape parameter is $\lambda > 0$ and in the case of power divergence loss, $\lambda < 0$. Consider the simplest scenario where only one observation is being fit and hence we can plot $L(\Delta)$ vs $\lambda$ over a sequence of $\lambda$ values for both loss functions, as in Figure ~\ref{fig:loss}. For the LINEX loss, the three lines represent $\Delta = \{-2,-3,-4\}$. Observe that minimum loss is always near 0 indicating that there is no benefit in a different estimate. Similarly, for the PDL loss, the three lines indicate three different estimates $\hat Y = \{5,4,2\}$ for the same true $Y = 10$ and it seems that all three estimates give minimum loss at the smallest $\lambda = -20$. This is inconclusive and is simply the nature of the function.

\begin{figure}
\subfigure[]
{\includegraphics[width = 0.5\textwidth]{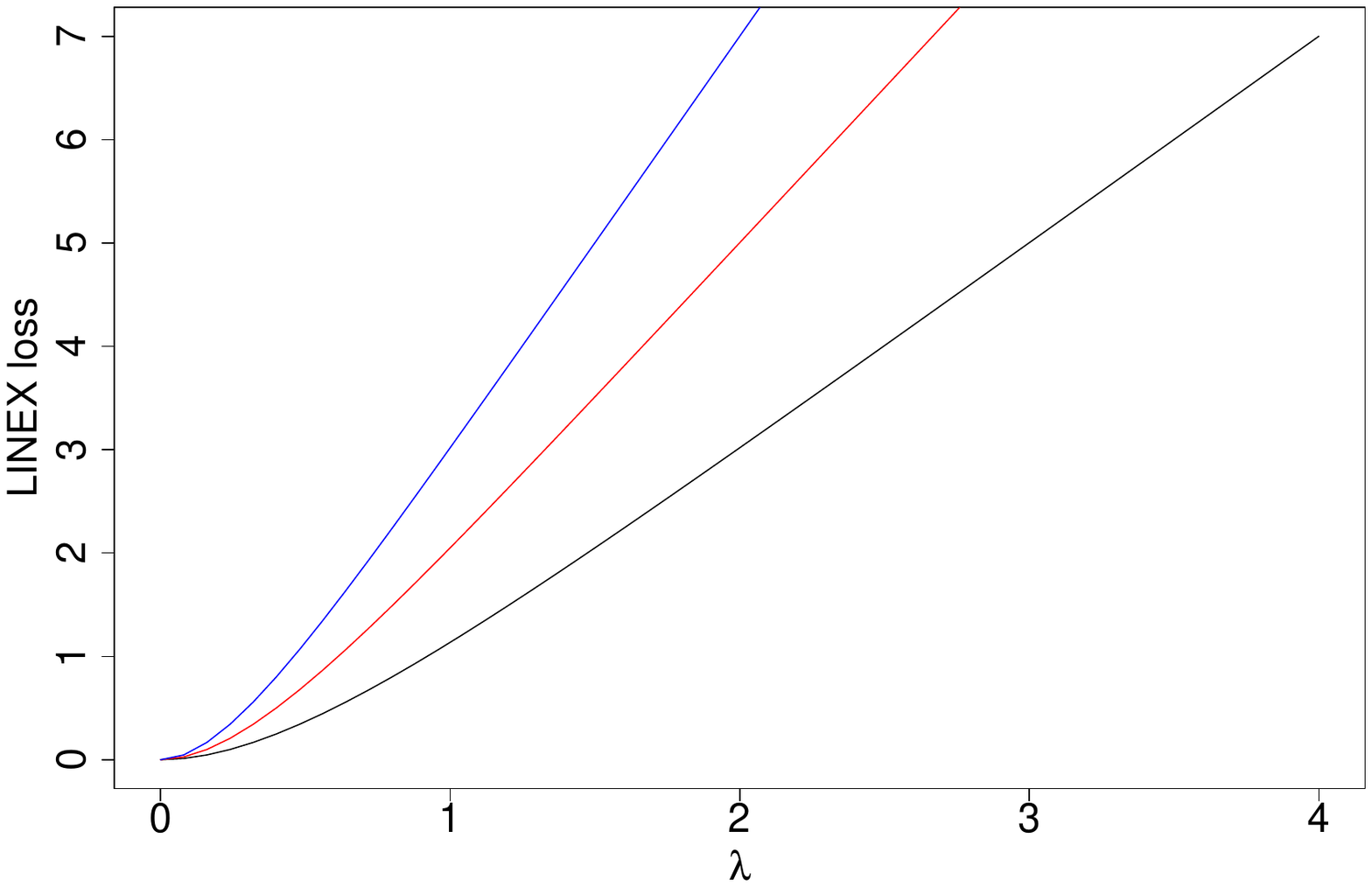}}
\subfigure[]
{\includegraphics[width = 0.5\textwidth]{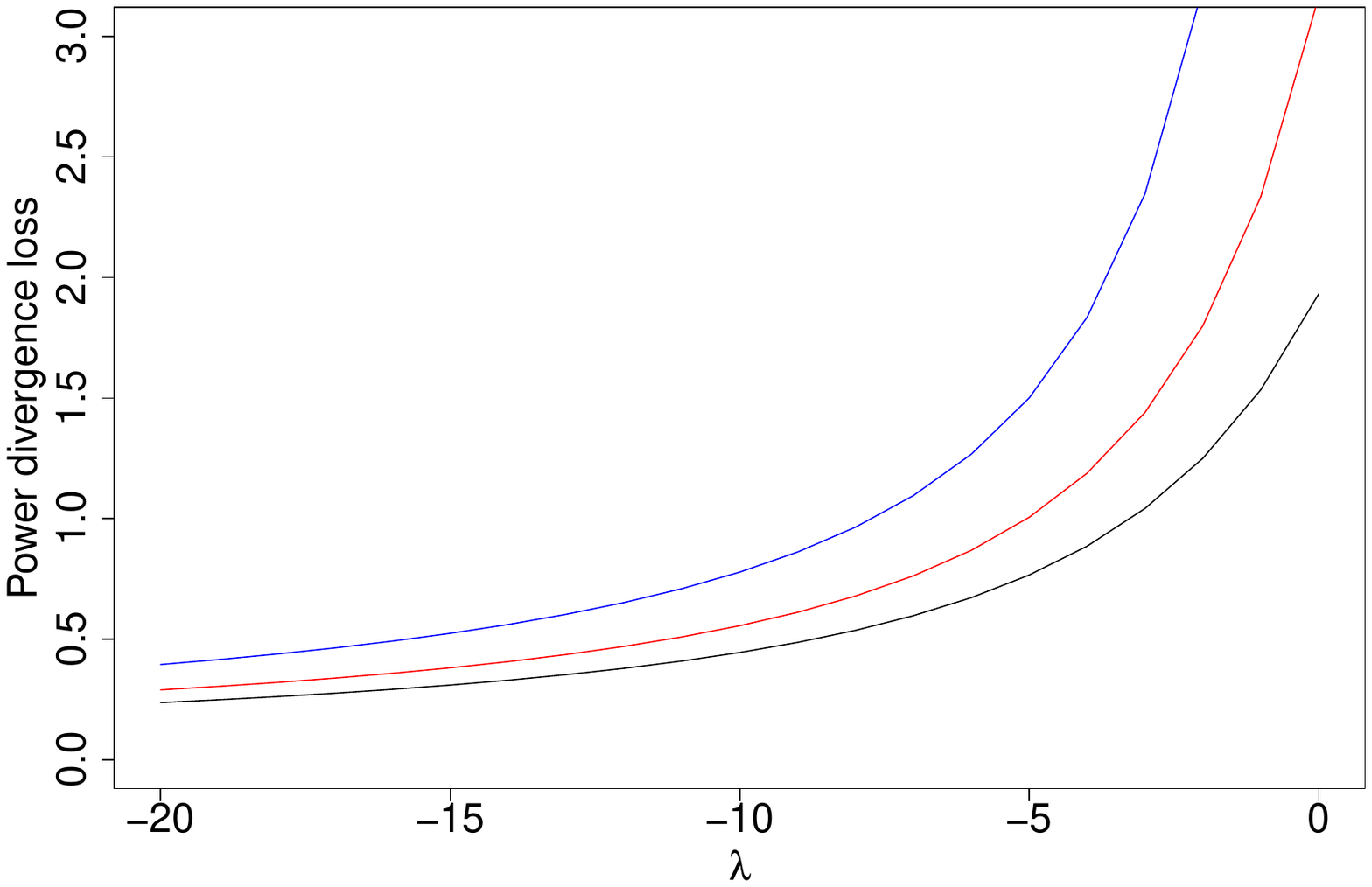}}
\caption{(a) Plot of LINEX loss in \eqref{eq:linexloss} versus asymmetry parameter $\lambda$ with 
$\Delta$ fixed as $\Delta = -2$ (black), $\Delta = -3$ (red), $\Delta = -4$ (blue) and (b) Plot of PDL in \eqref{eq:pdlloss} versus asymmetry parameter $\lambda$ with $\Delta = \hat Y - 10$ and $\hat Y=5$ (black), $\hat Y=4$ (red), $\hat Y = 2$ (blue).}
\label{fig:loss}
\end{figure}

As far as we know, there is no general recommendation for choosing the asymmetry parameter for these asymmteric loss functions. Here, we focus on the approach by \citet{jayasinghe2022regression}. The main idea being to penalize under (over) estimation more than the other, while guarding oneself against going more in the other direction than required. For this we define the power ratio $\Psi$
\begin{equation}
\label{eq:powerratio}
    \Psi (\lambda) = (RMSE^{+} \times R^{+})^{R^-} \times (RMSE^{-} \times R^{-})^{R^+},
\end{equation}
where for a residual as $\hat Y_i - Y_i$, $R^{+}$ and $R^{-}$ are the ratio of positive and negative residuals out of $n$ respectively. $RMSE^{+}$ and $RMSE^{-}$ are the root mean squared error for the positive and negative residuals respectively. The function $\Psi (\lambda)$ is plotted against $\lambda$ and the value of $\lambda$ that corresponds to a step/elbow in the slope of the curve is the appropriate choice. This is the point where the goal of guarding against under (over) estimation is achieved, while nothing significant will be achieved by increasing (decreasing) $\lambda$ further.

\end{itemize}

In the next section, we implement this methodology on the real estate dataset from Section \ref{s:dataset} to obtain real estate predictions based on asymmetric loss and compare the corresponding risks for both the predictors with the squared error loss predictor.  
\section{Analysis of Zillow Home Value Index (ZHVI)}
\label{s:analysis}
Zillow does not provide a measurement error associated with each ZHVI value. Hence, in this case we fit the CAR model without any measurement error. As described in Section~\ref{s:dataset}, we include the three covariates in-flow of migration, number of construction permits and county-level median household income in the mean of the process. The adjacency matrix $\bm C$ is constructed using a first-order nearest neighbor structure for the state of New Jersey with the boundaries for all $n=21$ counties such that $c_{ij} = 1$ if $j$ is in the neighborhood of $i$ and $0$ otherwise. The covariance matrix $\bm M$ is taken to be $\tau^2 \bm I_n$. The CAR model can then be written as
\[\bm Y \sim \nm (\bm{\mu}, \tau^{2}[\mathbf{I} - \rho \mathbf{C}]^{-1}), \]
where $\bm \mu = \bm X^\intercal \bm \beta$ with matrix $\bm X$ comprising of covariates and $\bm \beta$ denoting the regression coeffcients. We use vague priors on each of the unknown parameters as
\begin{equation}
    \begin{aligned}
        \beta_0 &\sim \nm(0, \sigma^2),\\ 
        \beta_j &\sim \nm(0, \sigma_j^2), \quad \text{for}~ j=1,2,3,\\
        \rho &\sim \unif(-1,1),\\
        \tau &\sim \stt(\text{df}=N, \text{location}=0, \text{scale} = \sigma_\tau^2).
    \end{aligned}
\end{equation}

\begin{figure}
\centering
\subfigure[\centering ]
{\includegraphics[width = 0.4\textwidth]{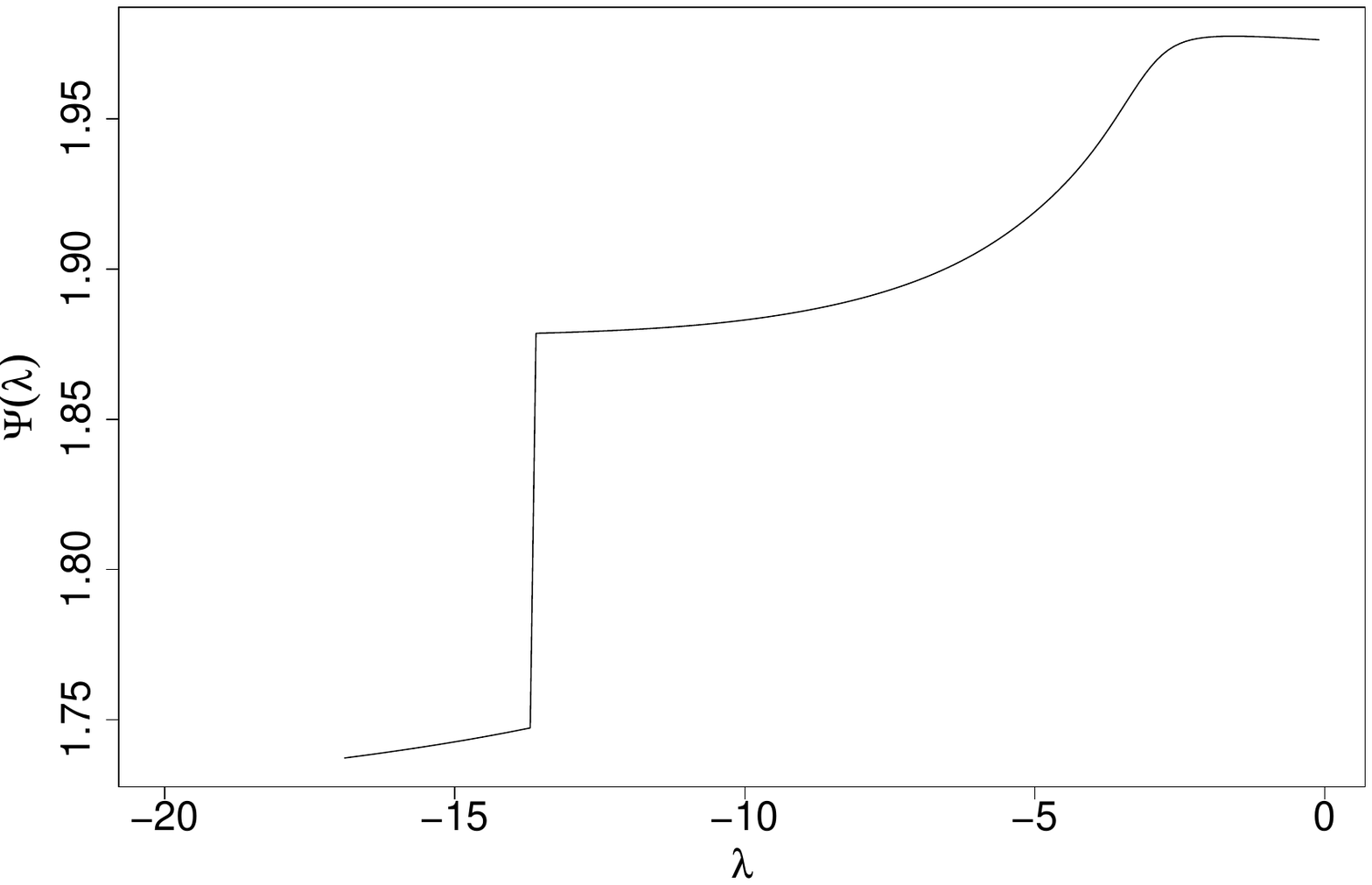}}
\subfigure[\centering ]
{\includegraphics[width = 0.4\textwidth]{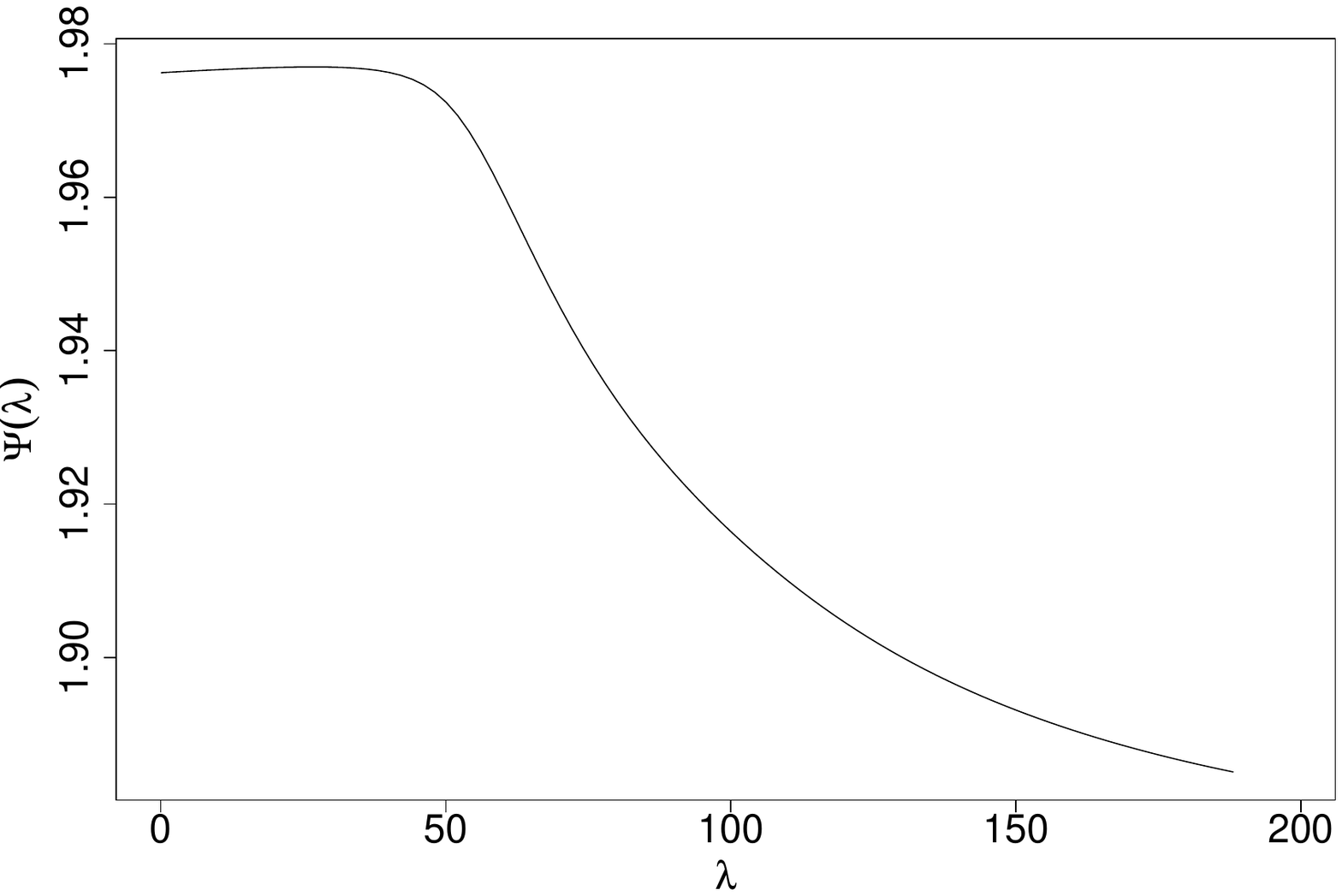}}
\subfigure[\centering ]
{\includegraphics[width = 0.4\textwidth]{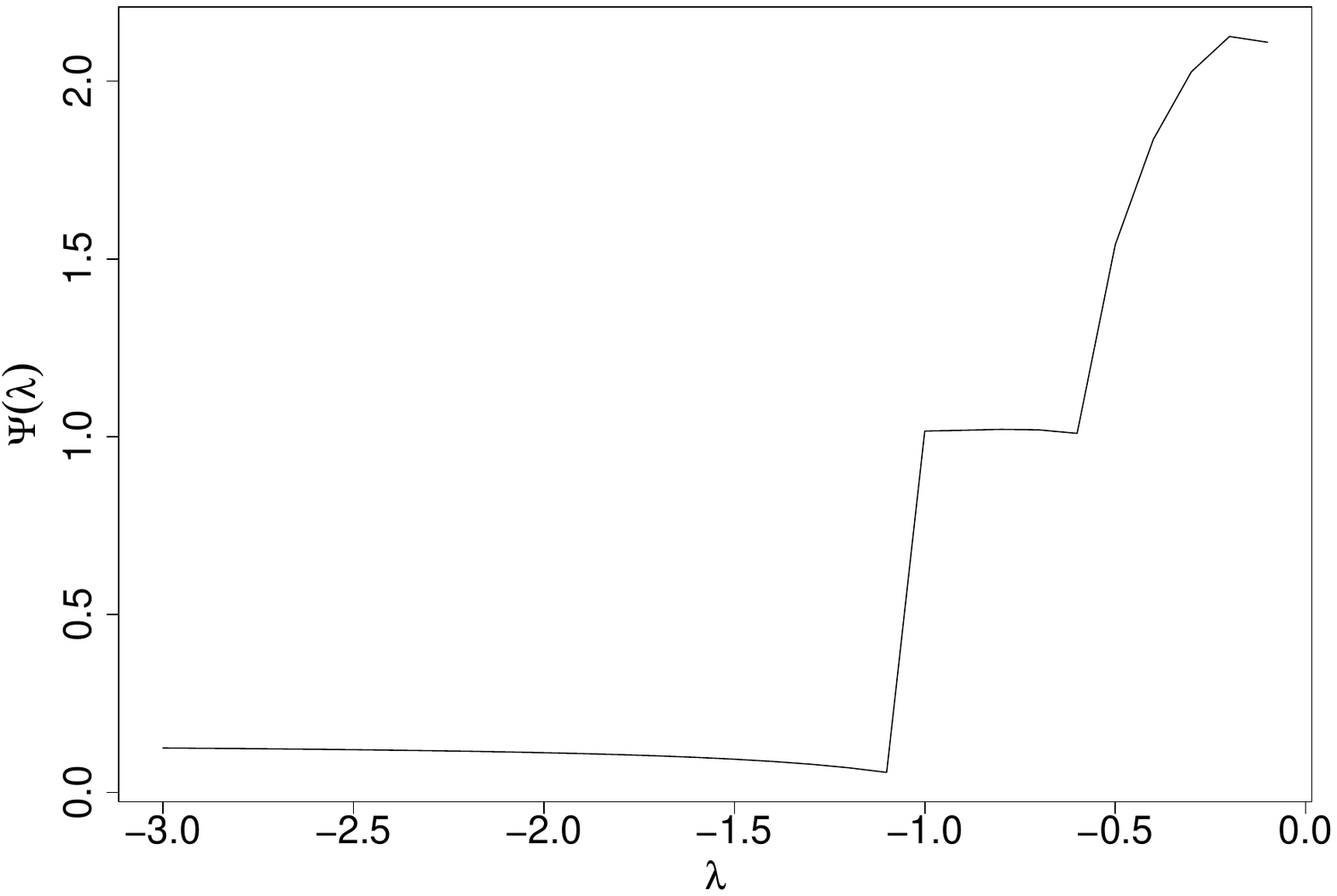}}
\subfigure[\centering ]
{\includegraphics[width = 0.42\textwidth]{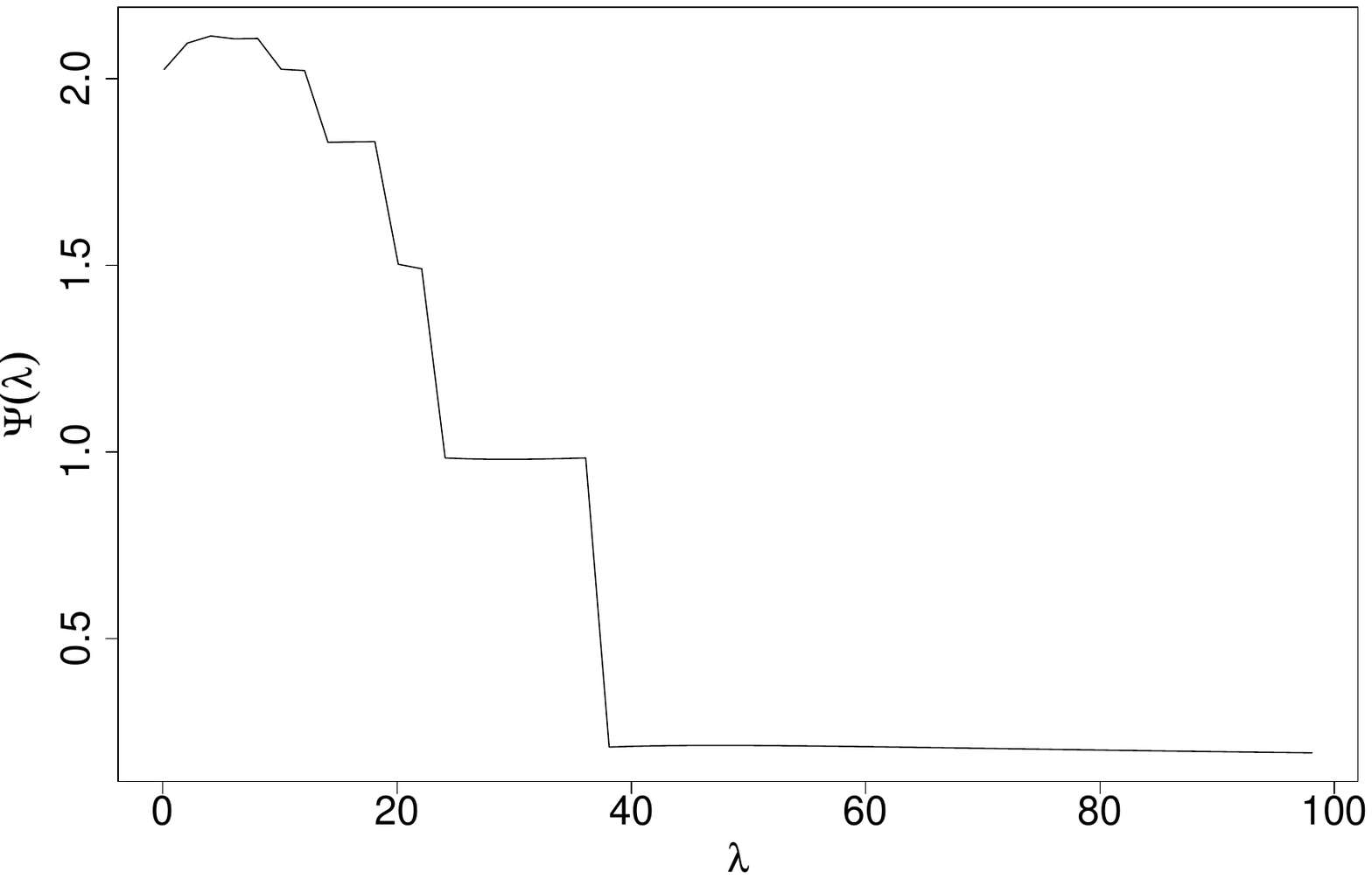}}
\caption{Plot of power ratio $\Psi(\lambda)$ given in \eqref{eq:powerratio} vs. asymmetry parameter $\lambda$ for the fitted values obtained from a CAR model without covariates (a, b) and with covariates (c, d), with LINEX loss (a,c) and PDL loss (b,d).}
\label{fig:powerratio}
\end{figure}

We run the model using the {\tt stan\_car} function in the R package {\tt geostan} with $M=15,000$ MCMC iterations including burn\-in iterations $=5000$ with hyperparameters as $\sigma^2 = 5, \sigma_j^2 = 5, \sigma_\tau^2 = 10, N = 15$. The package uses a dynamic Hamiltonian Monte Carlo algorithm to draw samples from the posterior distribution given the model specification above (see \citet{donegan2021building}). To compare a no-covariate model versus covariate model, we fit the CAR model under both scenarios. The kernel density estimates of the regression coefficients for the covariate model are symmetric and the corresponding $95\%$ credible intervals are $(-0.103, 0.349)$, $(-0.062, 0.100)$ and $(0.069, 0.296)$ for number of construction permits, number of movers, and median household income, respectively. The covariate of median household income shows significance. With the posterior draws from the fitted distribution, now the goal is to calculate the optimal predictions based on different loss functions. 
We focus from the seller's point of view that under estimation is costlier than over estimation, but the analysis for buyer's point of view is similar. For the LINEX loss and PDL function, we need to first choose the asymmetry parameter. For this we use the strategy in Section \ref{ss:asypar} and plot the power ratio $\Psi(\lambda)$ versus $\lambda$ for the appropriate range of $\lambda$, i.e., $\lambda > 0$ for the power divergence loss and $\lambda < 0$ for the LINEX loss. We plot these curves for both loss functions under both covariate and no-covariate models in Figure \ref{fig:powerratio}.

\begin{figure}
    \centering
    \includegraphics[width=0.7\textwidth]{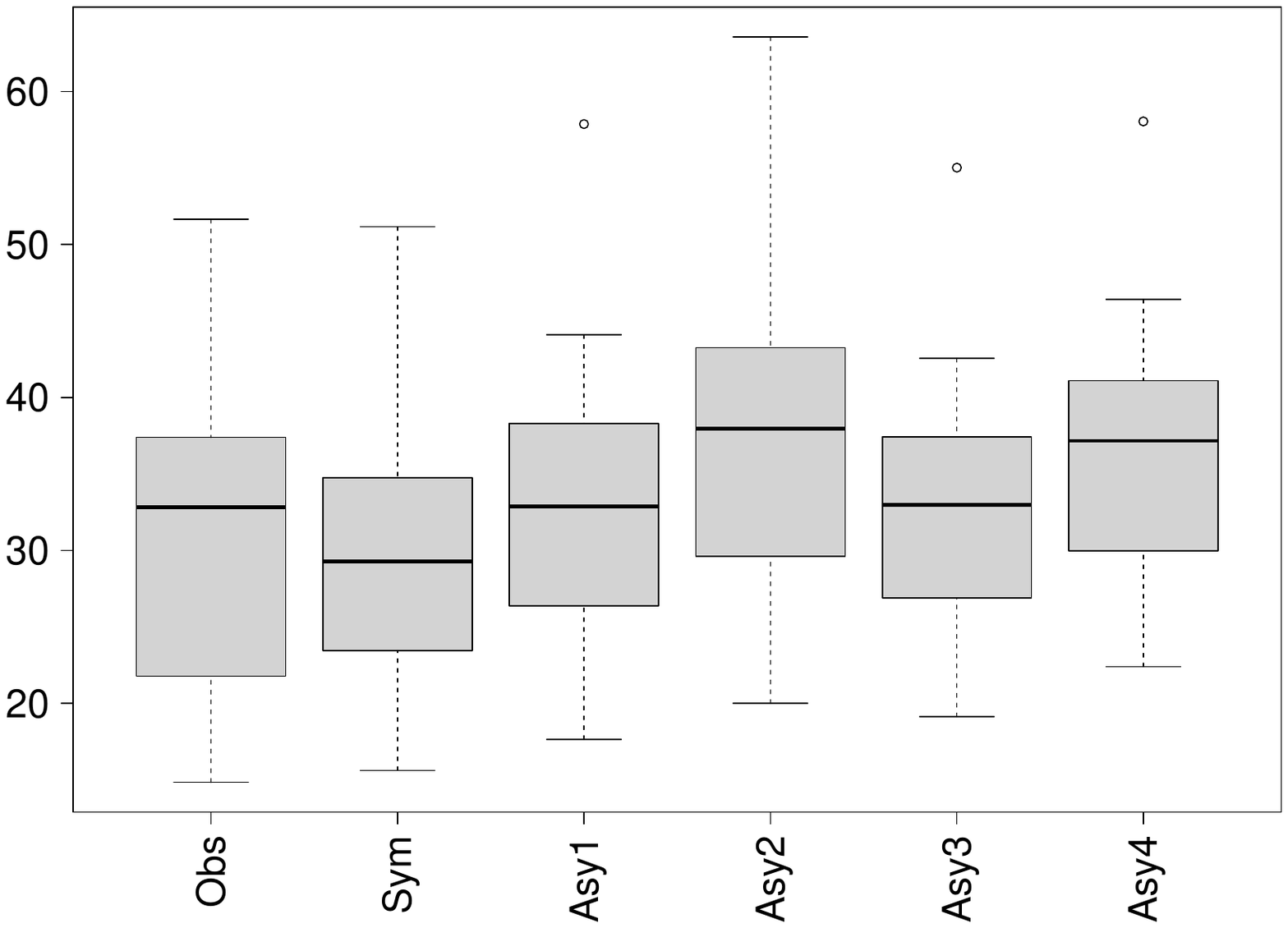}
    \caption{Boxplot of the observed and predicted ZHVI values with different loss functions. Sym is the posterior mean, Asy1 is the prediction under LINEX loss with $\lambda = -0.6$, Asy2 corresponds to LINEX loss with $\lambda = -1.1$, Asy3 is the PDL predictor with $\lambda = 22$ and Asy4 is the PDL predictor with $\lambda = 38$.}
    \label{fig:boxplot}
\end{figure}

\begin{figure}
\centering
\subfigure[]
{\includegraphics[width = 0.3\textwidth]{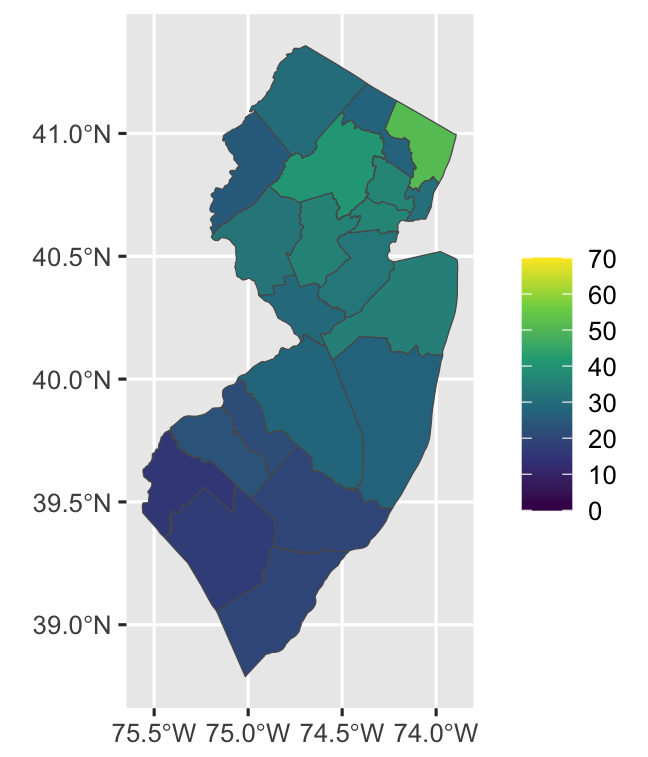}}
\subfigure[]
{\includegraphics[width = 0.3\textwidth]{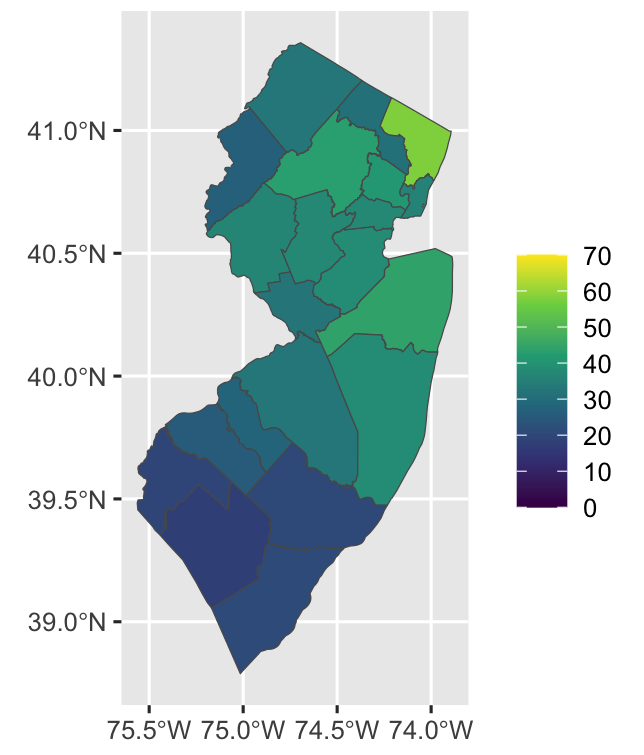}}
\subfigure[]
{\includegraphics[width = 0.3\textwidth]{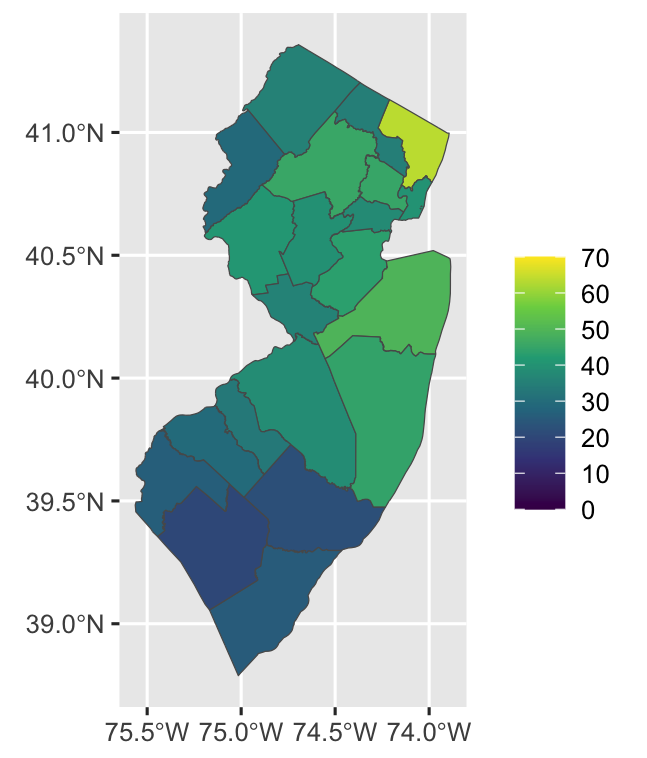}}
\subfigure[]
{\includegraphics[width = 0.31\textwidth]{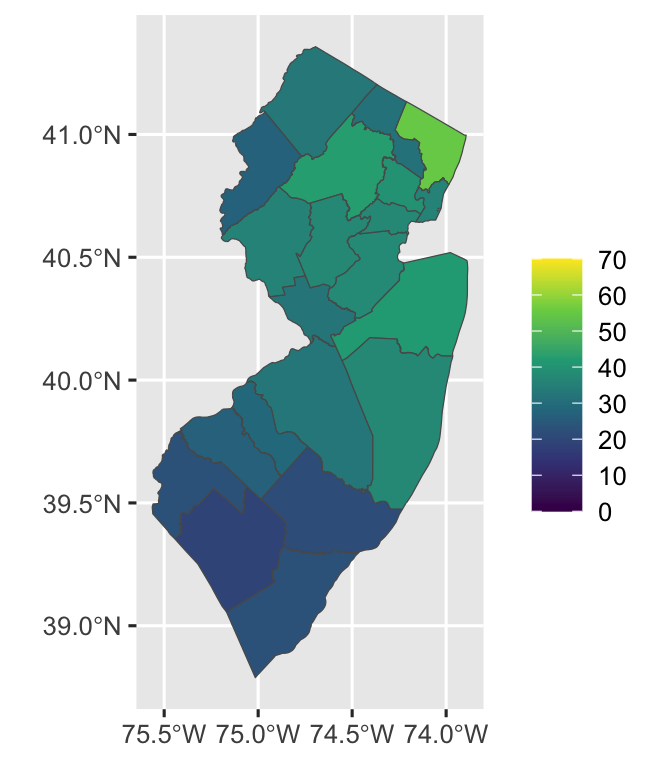}}
\subfigure[]
{\includegraphics[width = 0.3\textwidth]{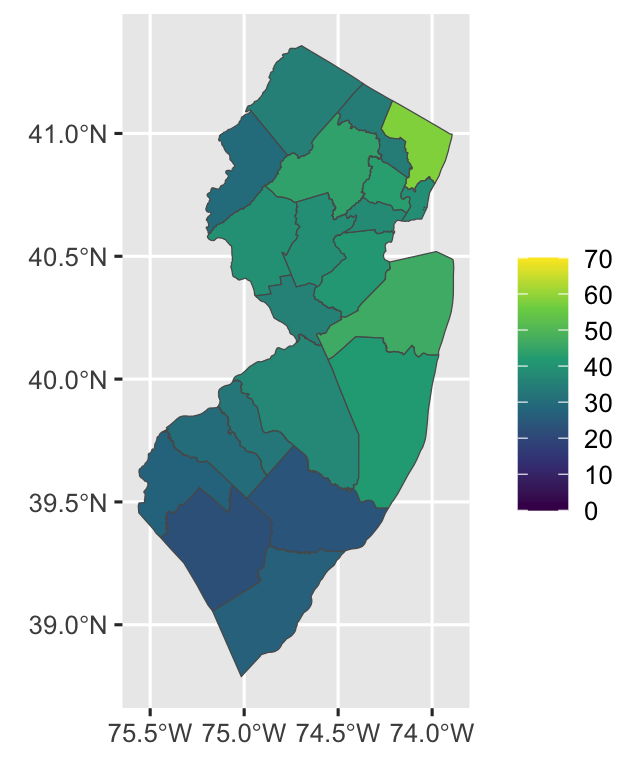}}
\caption{ Optimal prediction of the latent ZHVI under (a) squared error loss, (b) LINEX loss with $\lambda = -0.6$, (c) LINEX loss with $\lambda = -1.1$ (d) PDL with $\lambda = 22$ and (e) PDL with $\lambda = 38$.}
\label{fig:map_predictions}
\end{figure}

\begin{figure}
\centering
\subfigure[]
{\includegraphics[width = 0.3\textwidth]{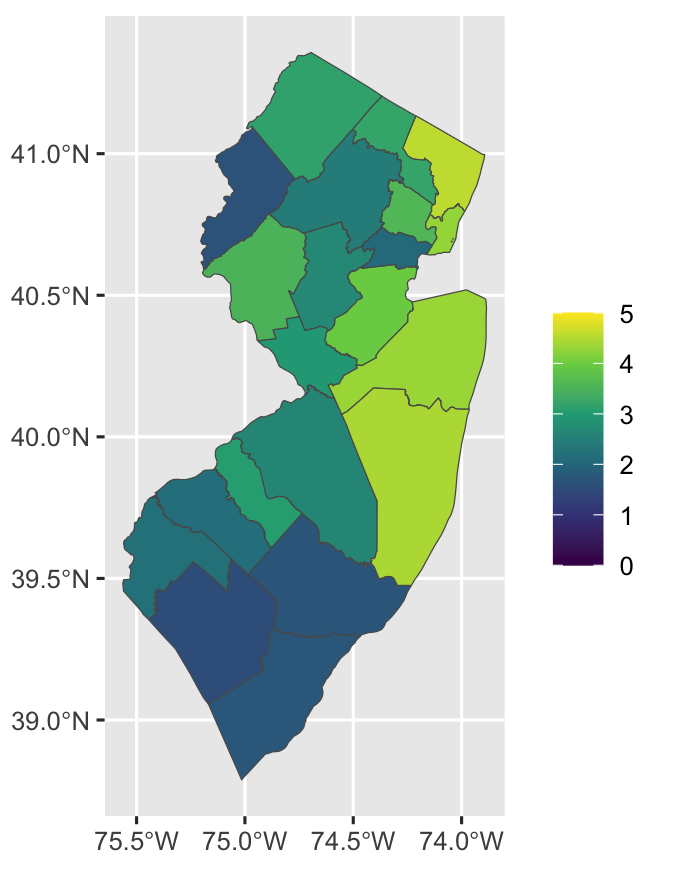}}
\subfigure[]
{\includegraphics[width = 0.31\textwidth]{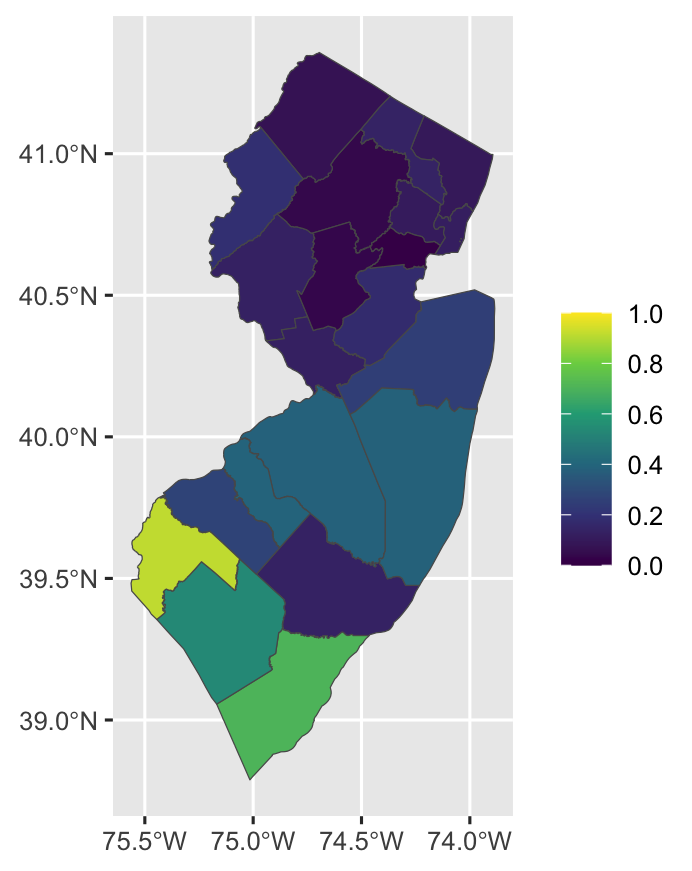}}
\subfigure[]
{\includegraphics[width = 0.3\textwidth]{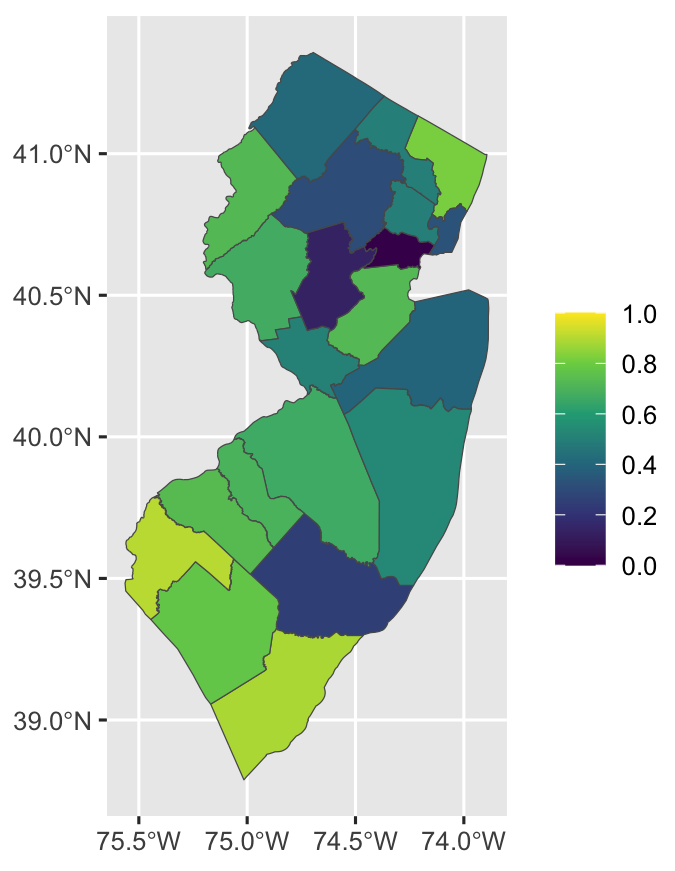}}
\subfigure[]
{\includegraphics[width = 0.31\textwidth]{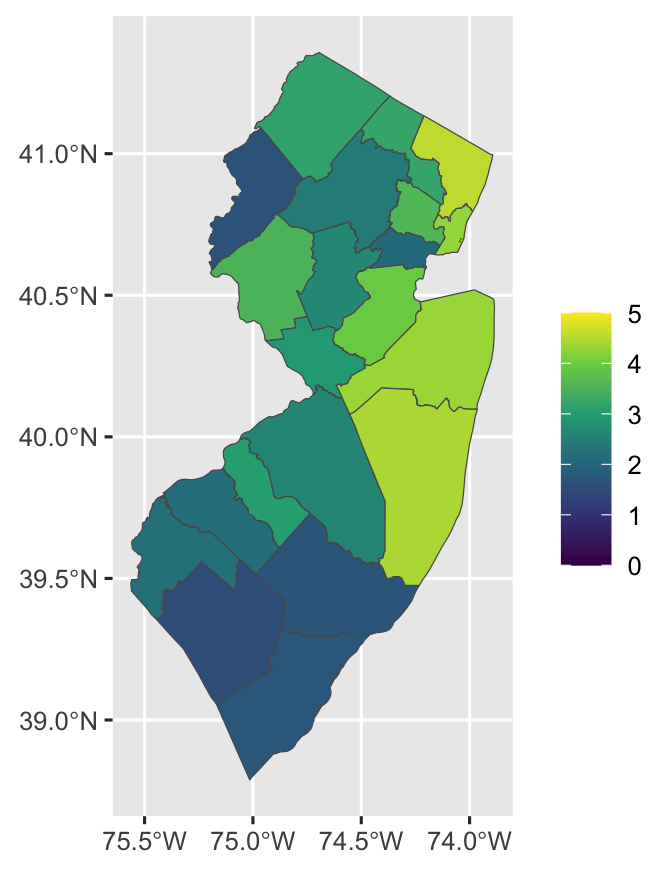}}
\subfigure[]
{\includegraphics[width = 0.31\textwidth]{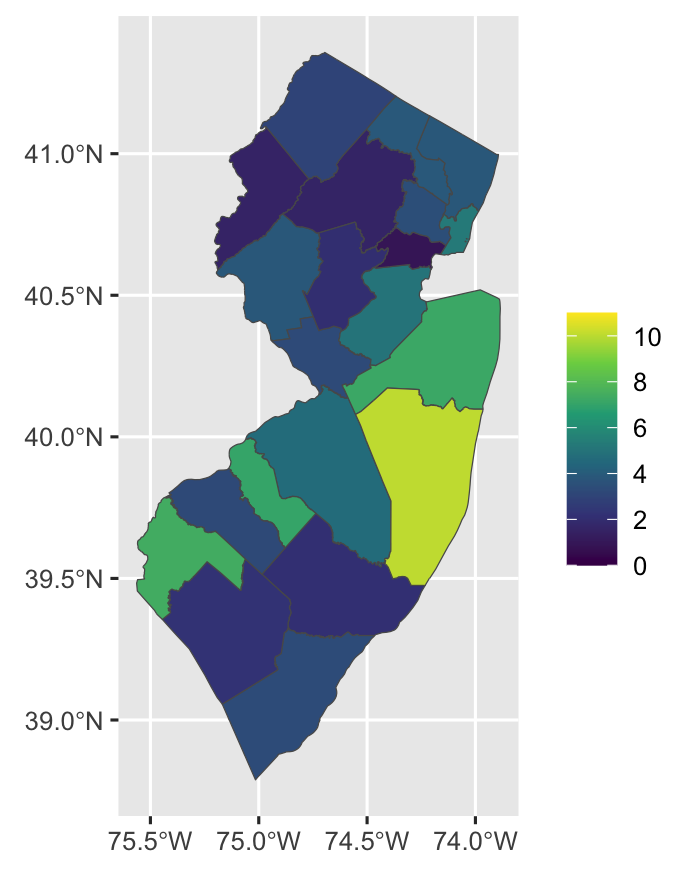}}
\subfigure[]
{\includegraphics[width = 0.31\textwidth]{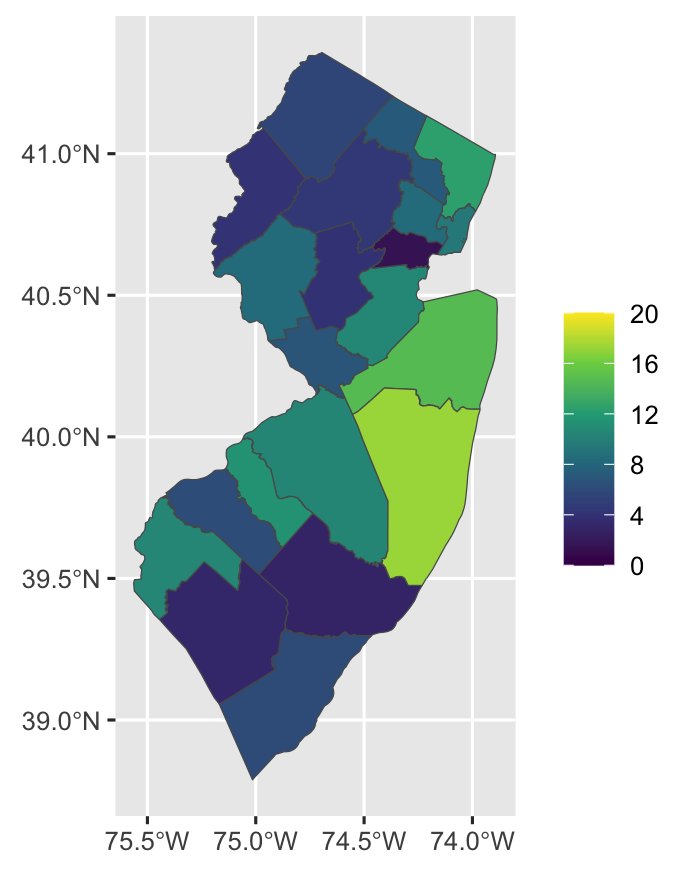}}
\caption{Standard deviation (a,b,c) and RMSPE (d,e,f) of the optimal predictors of ZHVI under squared error loss (a,d), PDL ($\lambda = 22$) (b,e) and LINEX ($\lambda = -1.1$) loss (c,f) respectively. Note the different range for each map.}
\label{fig:map_uncertainty}
\end{figure}

Under the no-covariate model a step in the slope is seen for a very large value of $\lambda$. This would lead to huge bias and overestimation. The situation is different for the model with covariates and we see that there are two possible choices $\lambda \in \{-0.6, -1.1\}$ under the LINEX loss and we can get the optimal predictor by \eqref{eq:linexestimate} under both. For the power divergence loss also, we can make a choice in $\lambda \in \{22,38\}$. We obtain the optimal predictor under each of these $\lambda$ for all 21 counties. Boxplots of the optimal predictors under both loss functions along with the posterior mean (optimal under squared error loss) are given in Figure \ref{fig:boxplot}. The predictors under the asymmetric loss functions are biased compared to the posterior mean by design, as their expected value is optimal under the respective asymmetric loss. All predictors under the asymmetric loss functions have a median higher (as expected) than the posterior mean. Maps of these predictions are given in Figure \ref{fig:map_predictions}. Further, we present the uncertainty quantification of the predictions in terms of the standard deviation and root mean squared prediction error (RMSPE). The latter accounts for the bias resulting from the asymmetric loss predictors. These plots are in Figure \ref{fig:map_uncertainty}. We see that the standard deviation of the optimal predictors in smaller for the asymmetric loss predictors than the posterior mean. However, as the bias adds in the RMSPE, the latter is larger for the asymmetric predictors. For a good tradeoff, the PDL predictor with $\lambda = 22$ has a small standard deviation and a relatively small RMSPE too.

\begin{figure}
    \centering
    \includegraphics[width=0.7\textwidth]{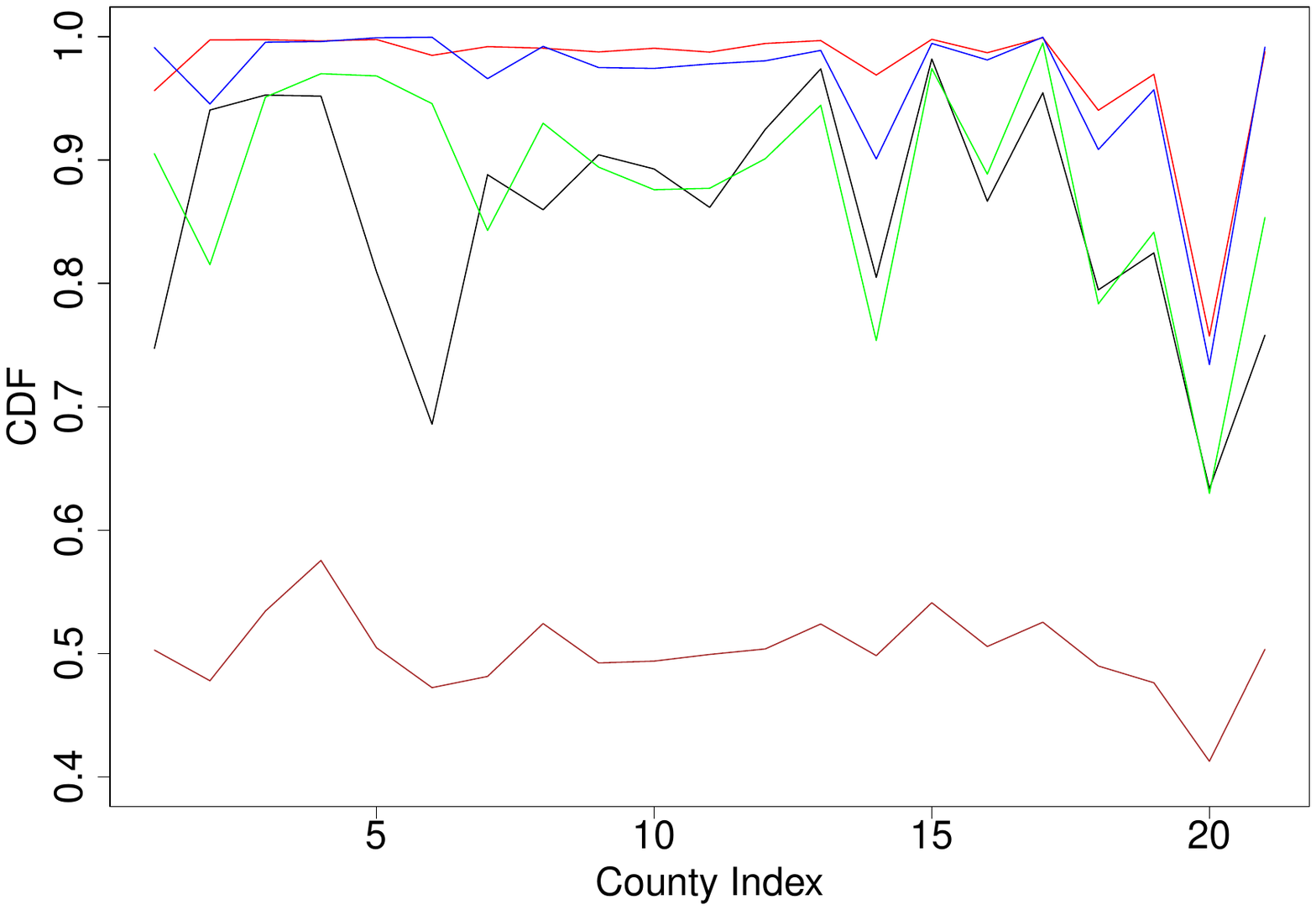}
    \caption{Plots of the quantile matched under different optimal predictors -- posterior mean (brown), LINEX loss with $\lambda = -0.6$ (black), LINEX loss with $\lambda = -1.1$ (red), PDL with $\lambda = 22$ (green) and PDL with $\lambda = 38$ (blue) for each county, for e.g., the LINEX loss optimal predictor for county index 1 corresponds to the $75^{th}$ quantile of the posterior distribution.}
    \label{fig:cdf}
\end{figure}

To compare our predictors against the quantile-based strategy in \citet{pearse2024optimal}, we calculate the quantiles that correspond to our optimal predictors. Note that the posterior distribution for each county is different, so the optimal predictor under the same $\lambda$ for each county  will match a different quantile. Figure \ref{fig:cdf} gives a plot of the quantile achieved by the optimal predictors under both loss functions for each county. The PDL loss predictor with $\lambda=38$ closely matches with the LINEX loss optimal predictor with $\lambda = -1.1$ with an average percentile value of $0.99$, while the PDL loss with $\lambda = 22$ matches the LINEX predictor with $\lambda = -0.6$ with average percentile value $0.85$. 

\begin{figure}
\centering
\subfigure[]
{\includegraphics[width = 0.45\textwidth]{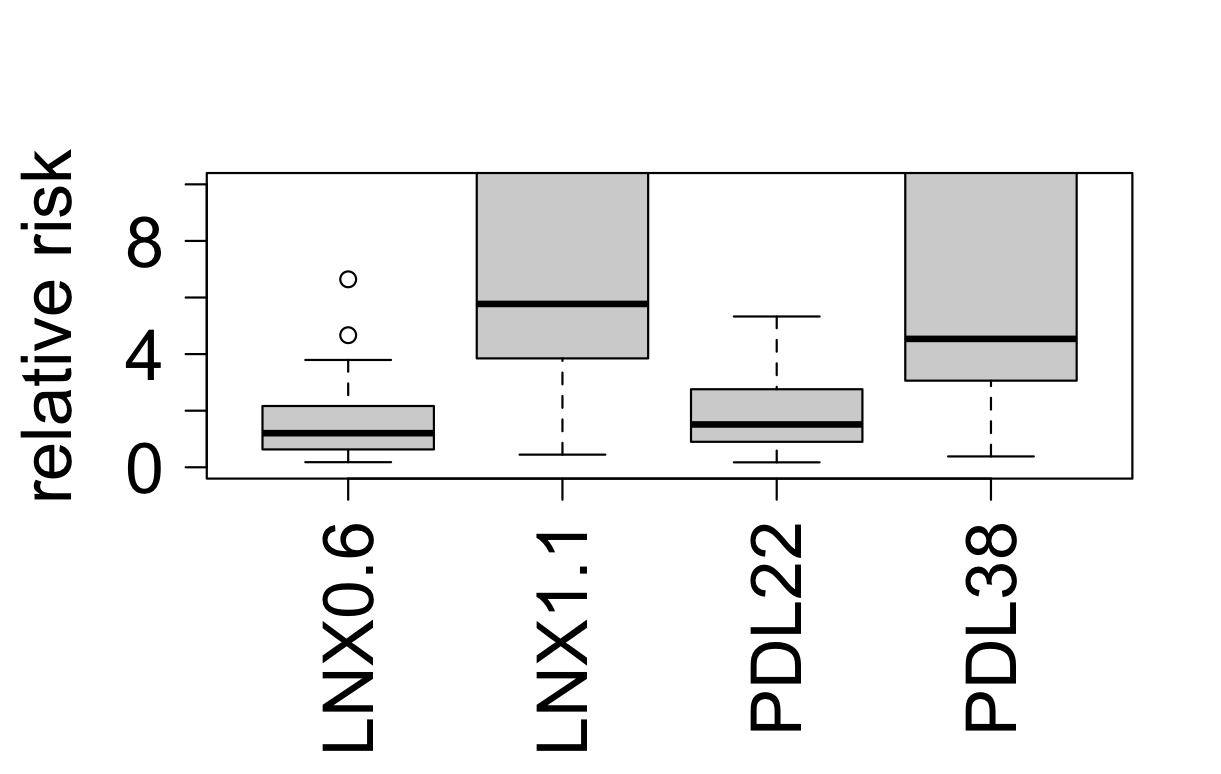}}
\subfigure[]
{\includegraphics[width = 0.45\textwidth]{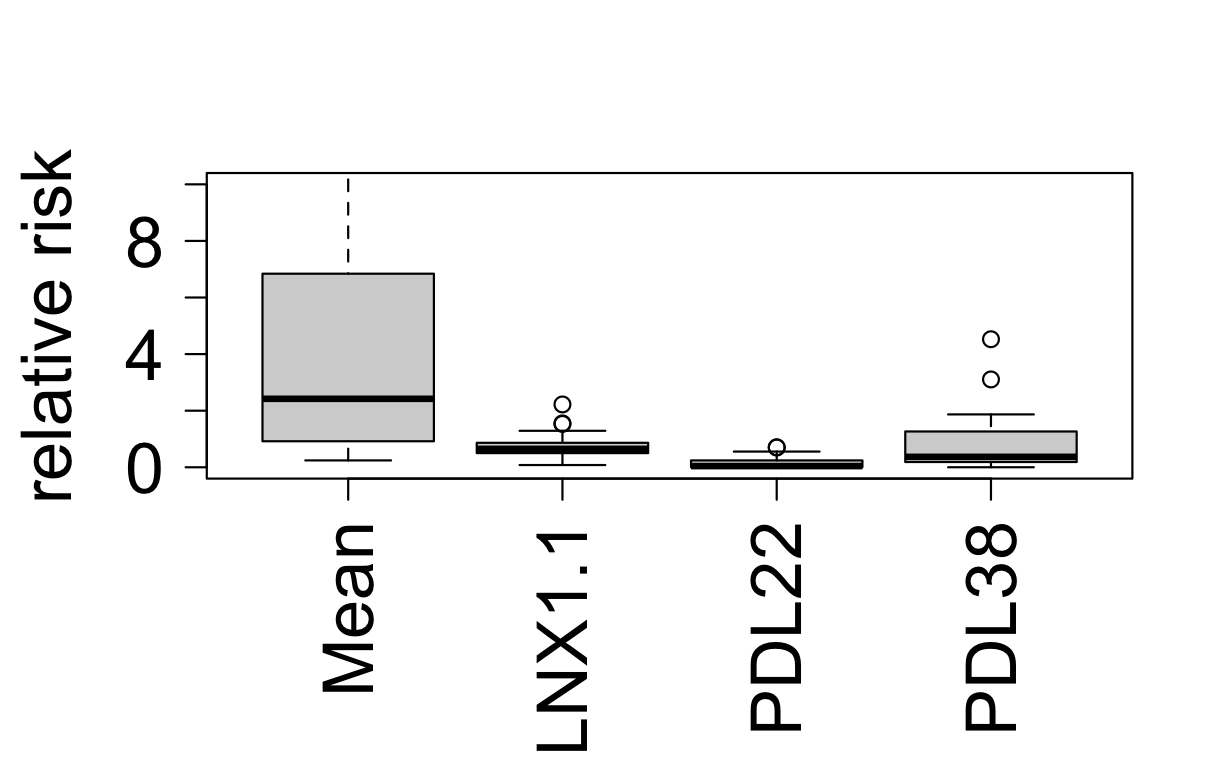}}
\subfigure[]
{\includegraphics[width = 0.45\textwidth]{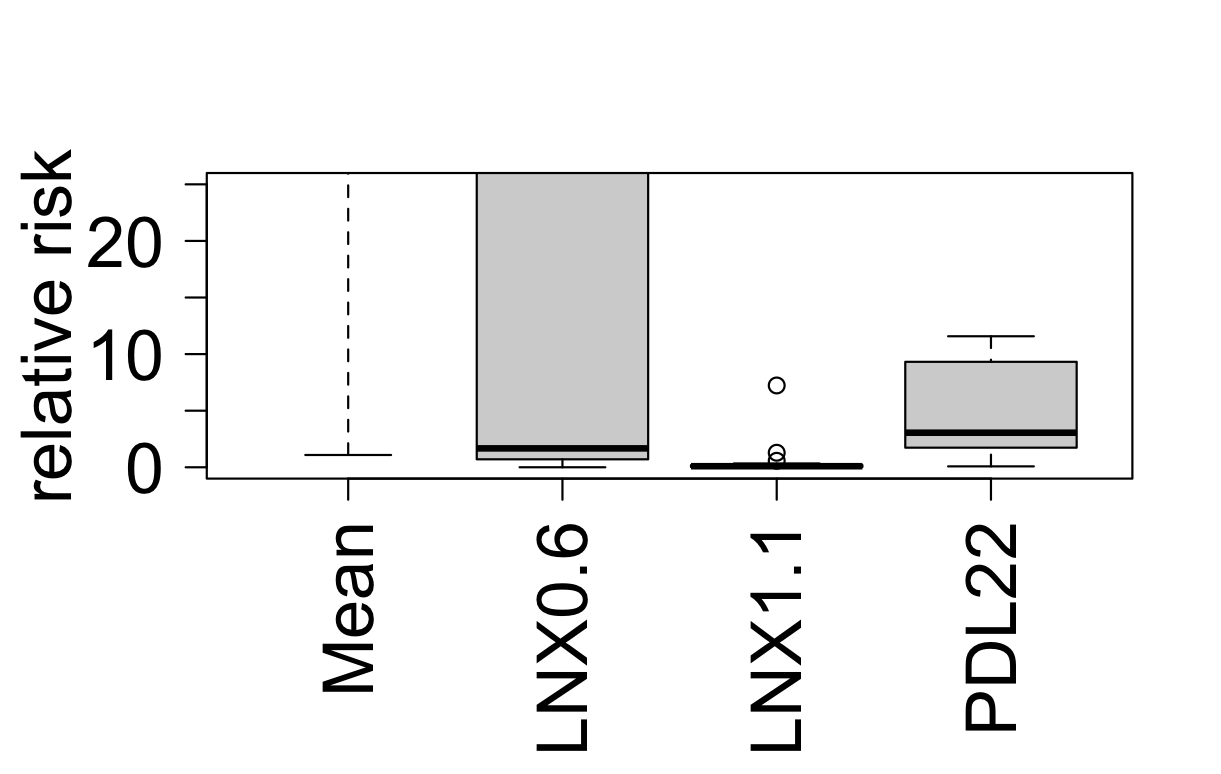}}
\caption{Boxplots of relative risk as in \eqref{eq:risk} for predictors as posterior mean (mean), LINEX loss with $\lambda = -0.6$ (LNX0.6), LINEX loss with $\lambda = -1.1$ (LNX1.1), PDL with $\lambda = 22$ (PDL22) and PDL with $\lambda = 38$ (PDL38) with respect to the corresponding optimal predictor under three true loss functions, namely (a) squared error loss (b) LINEX loss with $\lambda = -0.6$ and (c) PDL with $\lambda = 38$.}
\label{fig:risk}
\end{figure}

Since we are comparing different loss functions, we compare the risk of the predictors when the loss is misspecified, i.e., we know the risk will be minimized by the optimal predictor of the corresponding loss, but if in fact the predictor was incorrectly chosen, what is the degree of difference in risk? For this we define {\em relative risk} (RR) under true loss $l$ as
\begin{equation}
\label{eq:risk}
\text{RR}_i^l = \frac{\text{Risk}_l(\hat Y_i) - \text{Risk}_l(\hat Y_i^{OPT})}{\text{Risk}_l(\hat Y_i^{OPT})}, \quad i=1,\ldots,21,
\end{equation}
where for county $i$, $\text{Risk}_l(\hat Y_i)$ denotes the risk of predictor $\hat Y_i$ and $\text{Risk}_l(\hat Y_i^{OPT})$ denotes the risk of the optimal predictor $\hat Y_i^{OPT}$ under loss function $l$. We plot the relative risk under three loss functions, namely, squared error loss, LINEX loss with $\lambda=-0.6$ and PDL with $\lambda = 38$ in Figure \ref{fig:risk}. The inter-quartile range (IQR) for the same predictors under the above true loss functions is given in Table \ref{tab:IQR}. We can see that under the true loss as squared error loss (panel (a)) the relative risk is positive for all asymmetric predictors but is still smaller in magnitude for LINEX ($\lambda = -0.6$) and PDL ($\lambda = 22$) with the smallest IQR. The relative risk for the posterior mean is large under both asymmetric loss functions with a very high IQR, indicating that if we used the posterior mean as the predictor when in fact the true loss is asymmetric, then the risk is relatively high. This magnitude increases with increase in asymmetry in loss (from panel (b) to panel (c)). Interestingly, the IQR of predictor PDL22 under true loss LINEX ($\lambda = -0.6$) is the smallest and so is the IQR of predictor LNX1.1 under true loss PDL ($\lambda = 38$). We can conclude that optimal predictors under a flexible asymmetric loss would be able to capture the asymmetry and achieve the smallest risk even if the loss is in fact misspecified.

Overall, from the uncertainty quantification maps, the quantile matching and the relative risk (RR) calculations, the predictor under loss PDL ($\lambda = 22$) seems optimal as it has lower RR, is relatively less skewed and has smaller RMSPE relative to other asymmetric loss predictors. We can infer that this would be the most appropriate loss function to use in this scenario. In general, a user can implement this strategy to choose an appropriate asymmetric loss in other problems.

\begin{table}
\centering
\begin{tabular}{| *{5}{|c|}}
\hline
True loss & \multicolumn{4}{c|}{Predictor}  \\
\hline
Squared error loss & LNX0.6 & LNX1.1 & PDL22 & PDL38\\
IQR & 1.535 & 7.591 & 1.858 & 8.154\\
\hline 
LINEX $(\lambda=-0.6)$ & Mean & LNX1.1 & PDL22 & PDL38 \\
IQR & 5.921 & 0.364 & 0.239 & 1.080 \\
\hline 
PDL $(\lambda = 38)$ & Mean & LNX0.6 & LNX1.1 & PDL22\\
IQR & 4058.079 & 42.329 & 0.163 & 7.586\\
\hline 
\end{tabular} 
\caption{IQR of relative risk in eq.\eqref{eq:risk} of different predictors -- Mean (posterior mean), LNX0.6 (optimal under LINEX loss with $\lambda = -0.6$), LNX1.1 (optimal under LINEX loss with $\lambda = 1.1$), PDL22 (optimal under PDL with $\lambda = 22$ and PDL38 (optimal under PDL with $\lambda = 38$).} 
\label{tab:IQR}
\end{table}

\section{Discussion}
\label{s:discussion}
In this paper, we explored the implementation of two asymmetric loss functions, namely the LINEX loss and the power divergence loss function, in spatial prediction of area-level data. The flexibility and interpretation of the LINEX loss, still make it a very viable option for inclusion of asymmetric loss in a model. The power divergence loss gives the user the convenience to incorporate asymmetry for non-negative data and has similar flexibility as the LINEX. In our implementation with the CAR model, we showed that these loss functions can be included in spatial models with relative ease and improve the predictions, specially if the true cost of predictions is not symmetric. This can be an important tool specially in spatial literature as applications of the models range from environmental modeling and socio-economic analysis to commodity pricing across geographic areas. The need for incorporating asymmetry in our models is crucial when it comes to disease modeling, commodity valuation and climate change forecasts. Such an asymmetric loss can be included at different stages of analysis, in addition to the posterior phase as shown in this paper. For instance, the asymmetry in loss can be included for the regression coefficients when studying the relation of various covariates with the response at the spatial level.

We also focused on developing a criteria to choose the asymmetry parameter in an asymmetric loss in general. Combining developments across different methods, we suggest an approach that is informative and easy to use. The approach can be used to choose an asymmetry parameter for any asymmetric loss and can also be combined with other methods or domain knowledge to choose the best parameter value. Once the desired asymmetry is identified, optimal predictors under different asymmetric loss can be compared and the best one under the circumstance can be identified based on other diagnostic measures like uncertainty quantification or confidence intervals. In terms of future work, it would be desirable to develop a more objective solution or algorithm to choose an appropriate asymmetric loss given the data.

\section*{Acknowledgments}
  This article is released to inform interested parties of ongoing research and to encourage discussion. The views expressed on statistical issues are those of the authors and not those of the NSF or U.S. Census Bureau. This research was partially supported by the U.S. National Science Foundation (NSF) under NSF grant NCSE-2215168.

\bibliographystyle{apalike}

\end{document}